\documentclass[fleqn,usenatbib]{mnras}

\usepackage{newtxtext,newtxmath}

\usepackage[T1]{fontenc}

\DeclareRobustCommand{\VAN}[3]{#2}
\let\VANthebibliography\thebibliography
\def\thebibliography{\DeclareRobustCommand{\VAN}[3]{##3}\VANthebibliography}

\usepackage{graphicx}	
\usepackage{amsmath}	

\title[Detecting Ca II with Dual Neural Network]{Detecting Ca II Absorption Lines with a Fe II assisted Dual Neural Network}

\author[L. Wang et al.]{
Lucas Wang,$^{1,2}$\thanks{\href{mailto:lukewlh@gmail.com}{lukewlh@gmail.com} (LW)}
Jian Ge,$^{3}$\thanks{\href{mailto:jge@shao.ac.cn}{jge@shao.ac.cn} (JG)}
and Kevin Willis$^{1}$
\\
$^{1}$Science Talent Training Center, Gainesville, FL 32606, USA\\
$^{2}$Burnaby North Secondary School, 751 Hammarskjold Drive, Burnaby, BC V5B 4A1, Canada\\
$^{3}$Shanghai Astronomical Observatory, Chinese Academy of Sciences, Shanghai 200030, China
}

\date{Accepted XXX. Received YYY; in original form ZZZ}

\pubyear{\the\year{}}

\begin{document}
\label{firstpage}
\pagerange{\pageref{firstpage}--\pageref{lastpage}}
\maketitle

\begin{abstract}
Ca II absorbers, characterized by dusty and metal-rich environments, provide unique insights into the interstellar medium of galaxies. However, their rarity and weak absorption features have hindered comprehensive studies. In this work, we present a novel dual CNN approach to detect Ca II absorption systems, analyzing over 100,000 quasar spectra from the Sloan Digital Sky Survey (SDSS) Data Release 16. Our primary CNN identifies Ca II features, while a secondary CNN cross-verifies these detections using five Fe II absorption lines. This approach yielded 1,646 Ca II absorption systems, including 525 previously known absorbers and 1,121 new discoveries—nearly tripling the size of any previously reported catalog. Among our Ca II absorbers, 95 are found to show the 2175 \AA\ dust feature (2DA), corresponding to 22\% of strong absorbers, 7\% of weak absorbers, and $\sim$12\% of the overall Ca II population at $0.8 < z_{\text{abs}} < 1.4$. Across the full redshift range of $0.36 < z_{\text{abs}} < 1.4$, $\sim$1.5\% of Mg II absorbers host Ca II.
\end{abstract}



\section{Introduction}

The study of quasar absorption lines has significantly advanced our understanding of the interstellar medium (ISM), providing insight into the composition and evolution of galaxies across the universe. Among the many types of quasar absorption features, Magnesium II (Mg II) and damped Lyman-alpha (DLA) systems have been the primary focus due to their relative abundance and ease of detection in quasar spectra. 

DLAs, for example, are absorption systems with high H I column densities ($N > 2 \times 10^{20} \text{cm}^{-2}$; \citealt{1986ApJS...61..249W}). DLAs are great tracers for gas in the early universe, particularly warm and diffuse neutral gases \citep{2000A&A...364L..26P}. 

Calcium II (Ca II) absorption lines, particularly the H\&K doublet at rest-frame wavelengths of 3934 and 3969 \AA, offer critical insights into galactic environments, especially concerning dust and metal. Despite their importance, Ca II absorption lines are far less studied compared to Mg II or DLAs. One reason for this is the highly refractory nature of calcium, which leads to it being depleted in dust \citep{1996ARA&A..34..279S, 2005MNRAS.361L..30W, 2006MNRAS.367..211W, 2014MNRAS.444.1747S}. The low occurrence of Ca II absorption also results from its susceptibility to ionization. With an ionization potential of 11.873 eV—lower than neutral hydrogen's 13.598 eV \citep{NIST_ASD}—Ca II is not effectively shielded from ionizing radiation in neutral hydrogen environments. As a result, Ca II absorbers are rare, constituting only a small fraction of quasar absorption line systems. \cite{2014MNRAS.444.1747S} found that among the much more common Mg II absorption lines, only 3\% of them contain Ca II absorption lines. Previous studies \citep{2005MNRAS.361L..30W, 2006MNRAS.367..211W,2007MNRAS.374..292W, 2007MNRAS.379.1409Z, 2008MNRAS.390.1670N} have also shown that these absorbers typically reside in dense, dusty regions associated with significant star formation activity. They are often found in environments rich in metals and hydrogen, making them key tracers of the conditions favorable for star formation.

The dichotomy between strong ($W_0^{\lambda3934} \geq 0.7$ Å) and weak ($W_0^{\lambda3934} < 0.7$ Å) Ca II absorbers has been well-documented in prior research. Strong absorbers are commonly linked to low-impact parameter environments and disk-like galactic regions, whereas weak absorbers are more loosely linked with the gaseous halos of galaxies \citep{2006MNRAS.367..211W, 2014MNRAS.444.1747S, 2023MNRAS.518.5590F}. This division not only provides a framework for understanding the physical conditions of Ca II environments but also emphasizes their significance as complementary probes to Mg II and DLA systems. Strong absorbers, in particular, are noted for their high dust content and reddening, making them among the dustiest quasar absorbers known. \cite{2023MNRAS.518.5590F} identified and analyzed a set of 165 Ca II absorbers. Among this data set, they found that 33\% of strong Ca II absorbers exhibited the 2175 \AA\ dust bump (2DA), while only 6\% of weak absorbers showed this feature. This finding further reinforces the theory that strong absorbers predominantly reside in dusty, disk-like regions.

However, the limited amount of known Ca II absorbers currently prevents researchers from validating theories and models regarding these absorption lines. Traditional methods for identifying Ca II absorbers in large data sets often involve manual inspection and computationally expensive fitting procedures, making the process slow and labor-intensive. For example, \cite{2014MNRAS.444.1747S} adopted a process for searching Mg II absorbers originally developed by \cite{2005ApJ...628..637N}, \cite{Rimoldini2007}, and \cite{2011AJ....141..137Q}. This method involves normalizing spectra and then applying a sliding window to perform a line-fitting search for potential Ca II doublets. The targets with signal-to-noise ratios (SNR) greater than $5.0$ and $2.5$ on the two lines are flagged as detections and manually reviewed one at a time. This process is highly inefficient, as continuum and gaussian fitting, as well as manually reviewing thousands of spectra, all take considerable amounts of time.

To overcome these challenges, there is a growing interest in applying machine learning techniques, particularly deep neural networks, to the problem of absorption line detection. Convolutional neural networks (CNN) have proven effective in various astronomical applications, including the identification of absorption lines. \cite{2019MNRAS.487..801Z} developed a neural network to identify Mg II absorption lines at $\lambda \lambda$ 2796, 2803 \AA\ in Sloan Digital Sky Survey (SDSS) quasar spectra. This demonstrated the potential of CNNs to accurately identify Mg II absorption lines with high efficiency, as their model achieved 94\% accuracy and was able to search through 50,000 in just 9 seconds. \cite{2022MNRAS.517.4902X} adapted this method for identifying Ca II absorbers, and their model achieved 96\% accuracy on previously discovered absorbers by \cite{2014MNRAS.444.1747S}. Deploying that model on the Seventh and Twelfth Data Release of the SDSS (SDSS-DR7, DR12), it confirmed 409 previously known absorbers and identified 399 new ones with SNRs of the two lines above 3 and 2.5, respectively. This represented a significant improvement in detection accuracy as well as efficiency. Recently, \cite{2025ApJS..276...37L} developed a ResNet-CBAM model to identify Ca II absorbers as well as estimate equivalent width and full width at half-maximum values. Applied to SDSS DR7 and DR12 Mg II catalogs, the model rediscovers 321 known absorbers and identifies 381 new candidates with SNR thresholds set at slightly lower values of 2.5 and 2.

In this study, we adapt and improve the deep learning algorithm developed by \cite{2022MNRAS.517.4902X}, applying it to search for new Ca II absorbers in SDSS \citep{2000AJ....120.1579Y} DR 16 \citep{2020ApJS..250....8L}. Our dual-CNN approach, which cross-verifies Ca II detections using Fe II absorption features, enhances the accuracy and reliability of identifying these rare absorbers. This innovative method significantly reduces false positives while maintaining a high detection rate, addressing the limitations of traditional approaches.

In Section~\ref{sec:Data_and_Methods}, the search catalog and data sets are introduced, as well as the innovative dual CNN approach. In Section ~\ref{sec:Data_Processing_and_Neural_Network_Training}, data processing and neural network training are described. Section ~\ref{sec:Results} covers the results and data analysis, and discussions and concluding remarks are in Sections~\ref{sec:Conclusions_and_Discussions}.

\section{Data and Methods}
\label{sec:Data_and_Methods}

\subsection{Search Catalog}

Our investigation focused on analysing high-redshift Ca II absorber systems using new quasar spectral data from the SDSS DR16 quasar catalog \citep{2020ApJS..250....8L}. Due to the inherent weakness of Ca II absorption lines, they can be fairly difficult to detect. Therefore, we limited our search to the DR16 Mg II absorption catalog provided by \cite{2021MNRAS.504...65A}. Our final search catalog, with redshifts of $0.36<z_{\text{abs}}<1.4$, consisted of approximately 110,000 targets.

\subsection{Data Set}

Despite recent expansions in the catalog of quasar spectra containing Ca II absorption lines, the data set remains limited for robust training of neural networks. The scarcity of these spectra necessitates the creation of artificial training data sets. Following the approach by \cite{2022MNRAS.517.4902X}, we generated our synthetic data set from the DR16 \citep{2020ApJS..250....8L} Baryon Oscillation Spectroscopic Survey (BOSS) spectrographs, employing a method that injects artificial Ca II absorption lines into authentic quasar spectra. This technique ensures that the artificial data closely mimic the characteristics observed in real quasar environments, thereby providing a realistic training framework for our neural network model.

\subsection{Ca Test Set}

\begin{figure*} 
 \includegraphics[width=\textwidth]{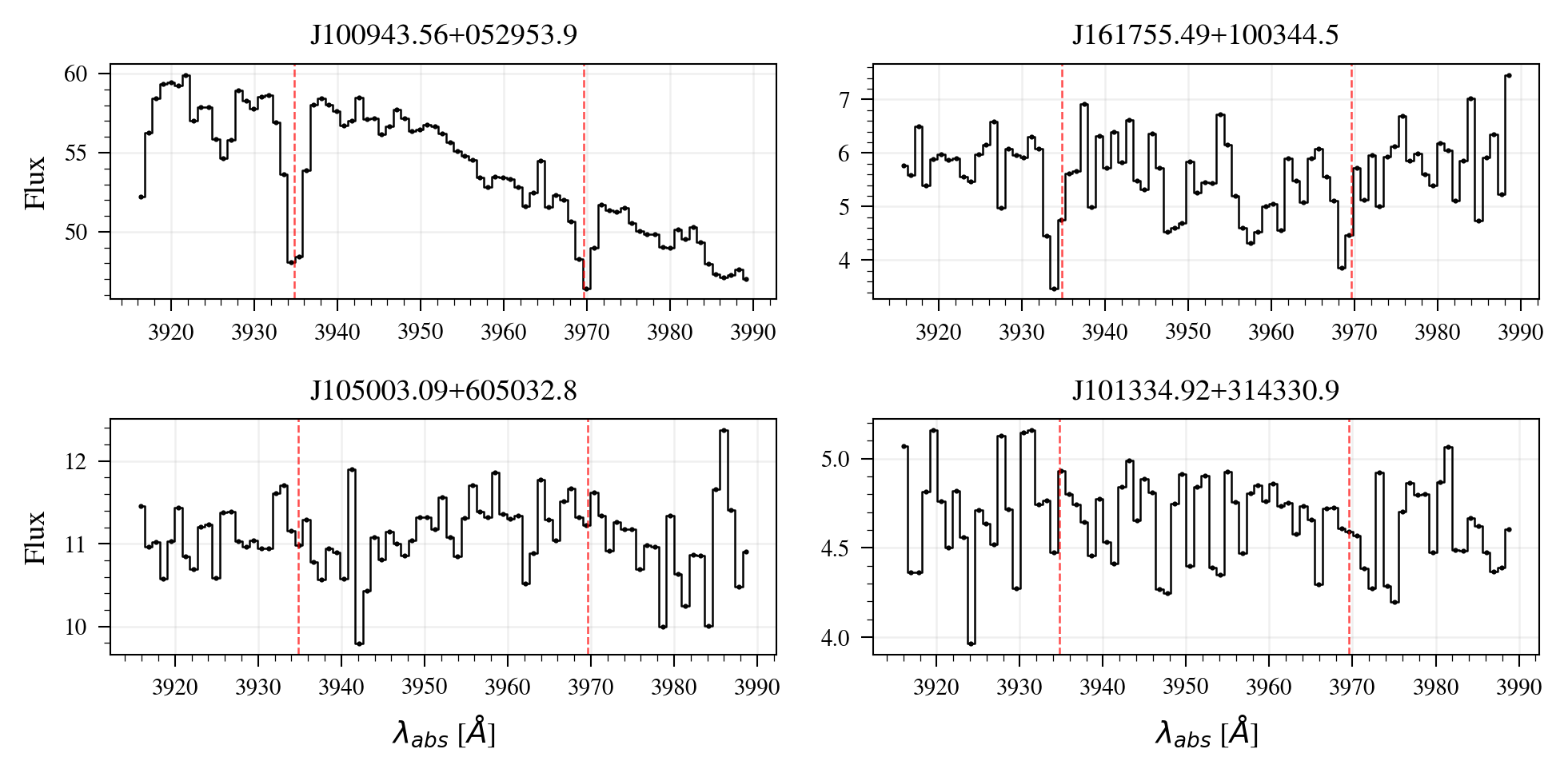} 
 \caption{Examples of quasar spectra from the Ca II test set. The red dotted lines mark the expected positions of the Ca II $\lambda\lambda$ 3934, 3969 absorption lines. The top two spectra are positive samples containing Ca II absorption features, taken from the catalog provided by \citet{2022MNRAS.517.4902X}. The bottom two spectra are false samples randomly selected from the Mg II absorption catalog provided by \citet{2019MNRAS.487..801Z}, and do not contain Ca II absorption.} 
 \label{fig:caii_testset_ex}
\end{figure*}

For the evaluation of our Ca II CNN, we utilized a modified subset of the quasar spectra data set identified by \cite{2022MNRAS.517.4902X}. Initial scrutiny of this data set revealed that 28 of the 399 newly identified spectra were duplicates among themselves, which reduced the number of unique spectra to 371. 

We then conducted a thorough fitting process of these 371 spectra, with a focus on verifying the presence of Ca II absorption lines. Spectra in which the fitted Gaussians failed to meet the SNR thresholds of 3.0 and 2.5 for the two lines \citep{2022MNRAS.517.4902X} were removed. Surprisingly, many detections did not show any absorption, so those spectra were excluded. Although some of the excluded targets lay just below the SNR thresholds, they were still removed for consistency.

This review process allowed us to identify 305 spectra with clear and well-defined Ca II absorption lines. These 305 spectra became the foundation of our test set, providing a wide variety of real-world conditions that were instrumental in assessing the performance and robustness of our neural network. The selection of these spectra ensured that our model was validated against actual observational data that included minimal amounts of false detections.

To balance the data set, we also randomly selected 305 false test samples to serve as the negative half of our test set. These negative samples were drawn from the same Mg II absorption catalog used in the initial search conducted by \cite{2022MNRAS.517.4902X}, and the spectral windows were cropped at the expected location of Ca II lines based on the Mg II absorption redshift. By using this approach, we ensured that the CNN was exposed to both true and false examples in a controlled and realistic manner, enabling us to rigorously evaluate the model's ability to accurately distinguish genuine Ca II absorption lines from noise or other spectral features. Some examples of spectra in the test set are shown in Figure~\ref{fig:caii_testset_ex}.

We applied a similar process to the Ca II catalog from \cite{2014MNRAS.444.1747S}, resulting in 412 confirmed absorbers. Lastly, we applied the same filtering process to the 381 new absorbers identified by \cite{2025ApJS..276...37L}. Many of these were excluded during verification, resulting in a final subset of 256 confirmed absorbers. A major reason for the large number of removals was the lower SNR thresholds used in their analysis—2.5 and 2.0 for the Ca II lines. To maintain consistency and reliability throughout this study, we adopted stricter thresholds of 3.0 and 2.5, and applied them uniformly across all catalogs.

\subsection{The Faked Quintuplet Method}

\begin{figure*} 
 \includegraphics[width=\textwidth]{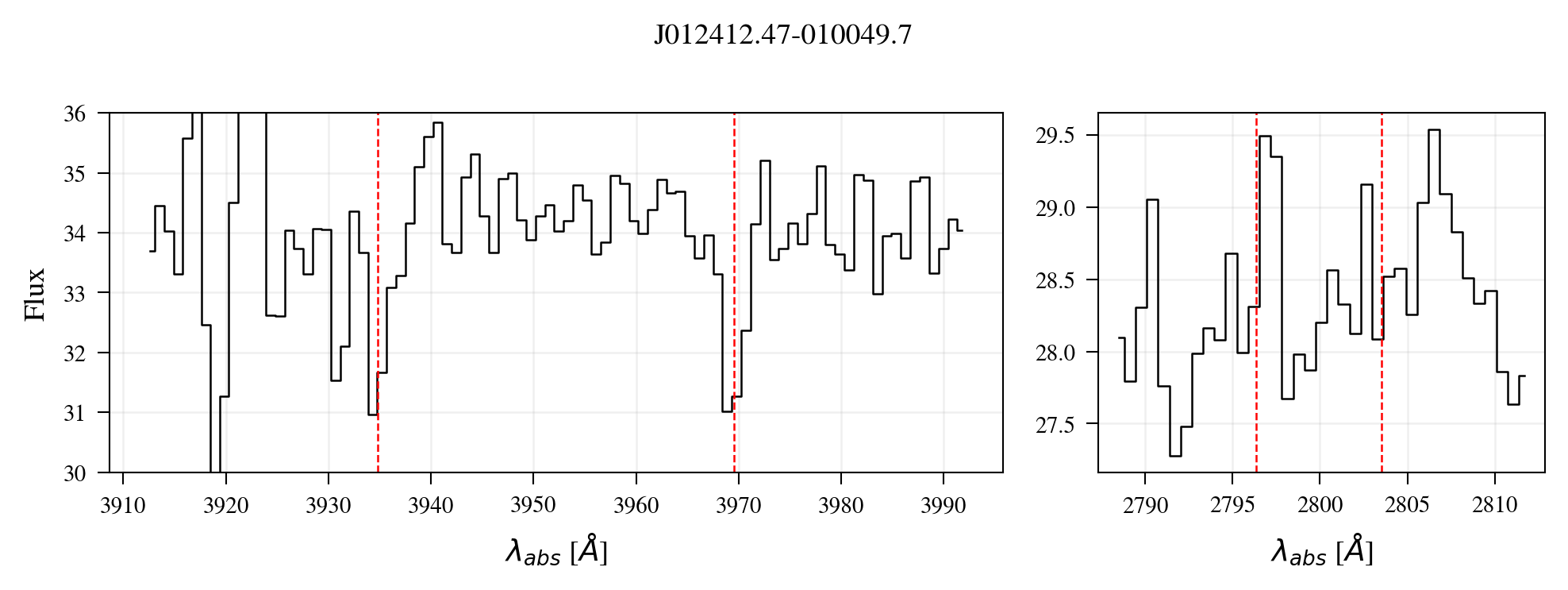} 
 \caption{Example of a Ca II absorber from \citet{2014MNRAS.444.1747S} that lacks visible Mg II absorption. The left panel shows the Ca II doublet, while the right panel displays the Mg II region, with red dashed lines indicating the expected line positions based on the absorption redshift. } 
 \label{fig:sardane_false}
\end{figure*}

In our analysis, the CNN trained to identify Ca II absorption lines produced false positives, which was a common challenge in neural network applications for spectral line detection. Previous studies, such as those by \cite{2024MNRAS.531..387G} and \cite{2022MNRAS.517.4902X}, have also encountered this issue. The high rate of false positives persists despite achieving high model accuracy, and it is partly due to the inherent weakness of the Ca II absorption lines. In certain cases, this makes them very difficult to differentiate from noise or other spectral features.

Given this ambiguity, we adopted an innovative dual-network approach to improve the accuracy of our detections and reduce the number of initial false positives. Specifically, we introduced a second CNN focused on identifying Fe II $\lambda \lambda$ 2344, 2374, 2382, 2586, 2600 absorption lines. These Fe II lines, typically stronger and more prominent than the H\&K Ca II lines, occur in similar physical environments, and their presence or absence can help confirm or refute the detection of Ca II. In cases where the Fe II CNN identified no significant absorption, the corresponding Ca II detection was rejected as a false positive. This dual CNN system cross-verified Ca II detections by checking for corresponding Fe II absorption, reduced the number of false positives, and maintained a high detection rate for true Ca II absorbers. 

To test that Fe II lines can indeed be used to verify Ca II lines, we analyzed the Ca II catalogs provided by \cite{2022MNRAS.517.4902X} and \cite{2014MNRAS.444.1747S}. Among the 305 detections in \cite{2022MNRAS.517.4902X}, 202 had redshifts such that all five Fe II lines were within the spectral range and outside the Lyman $\alpha$ forest. Manually analyzing each of these targets, only $2$ of them did not show observable Fe II lines. In \cite{2014MNRAS.444.1747S}, there were 131 targets within the 412 total targets that had redshifts in range. Among these, 13 targets did not have observable Fe II lines. However, 12 of these 13 spectra also did not have observable Mg II absorption lines, with one such a spectra is shown in Figure~\ref{fig:sardane_false}. Thus, we concluded that using Fe II lines to verify Ca II detections introduces no significant bias when searching through Mg II absorbers.

\section{Data Processing and Neural Network Training}
\label{sec:Data_Processing_and_Neural_Network_Training}

\subsection{Artificial Ca II Data Set}

\begin{figure*}
 \includegraphics[width=\textwidth]{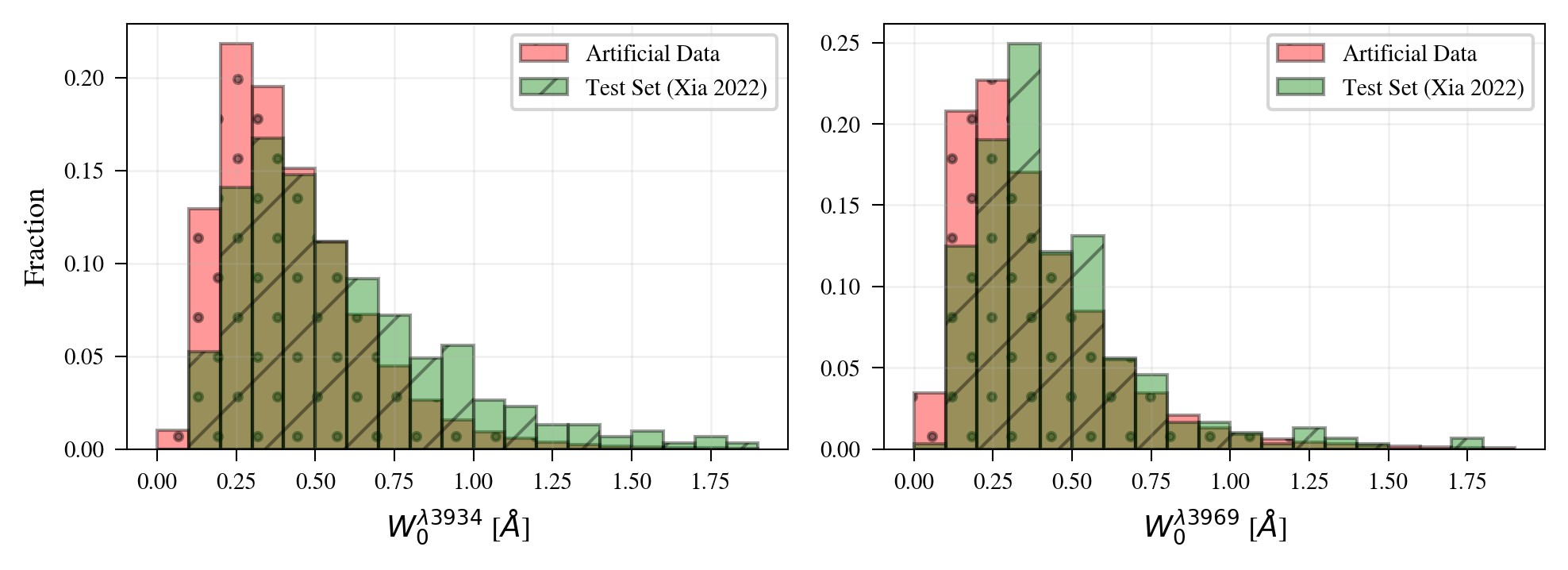}
 \caption{The distribution of the EW for the catalog by \citet{2022MNRAS.517.4902X} compared to the distribution of the measured EW of the our artificial training data. The training samples have slightly lower equivalent widths to improve our CNNs sensitivity.}
 \label{fig:EW_distributions}
\end{figure*}

\begin{figure}
 \includegraphics[width=\columnwidth]{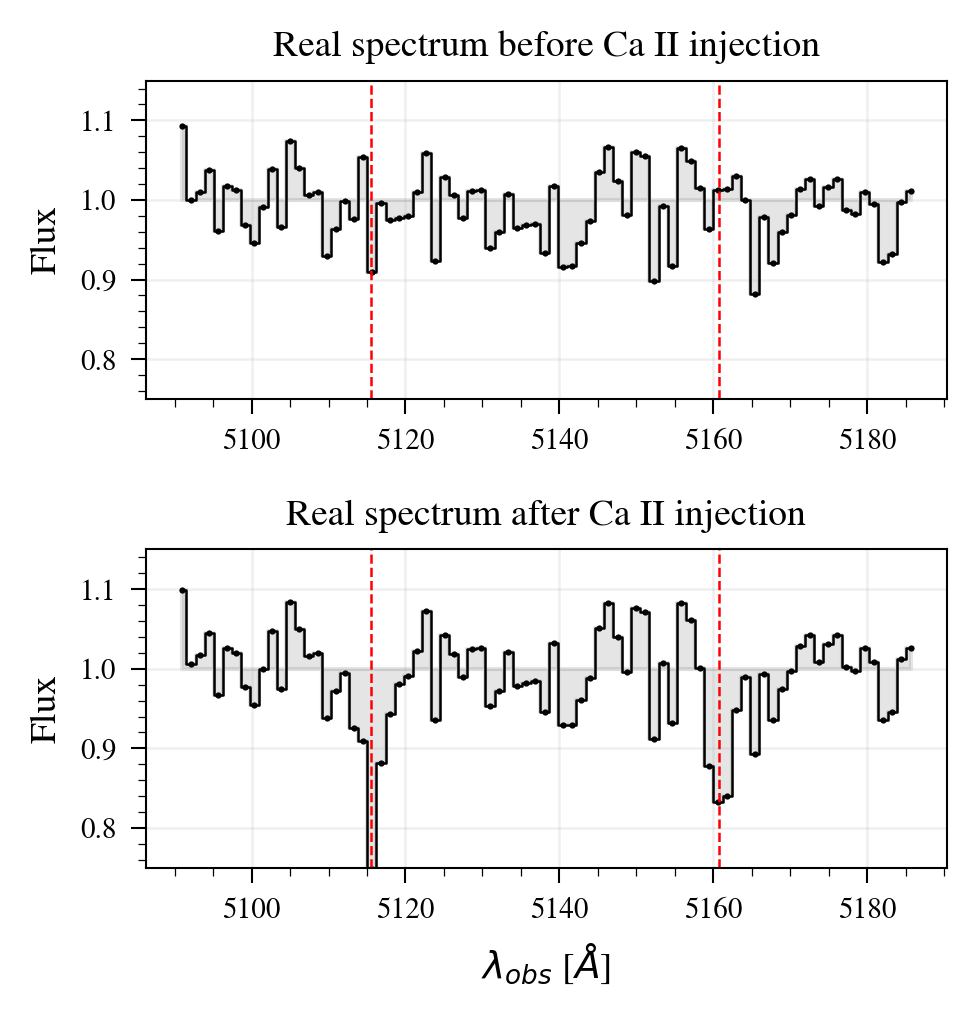}
 \caption{Samples of artificial spectra that are generated. The red dotted lines indicate the locations at which Ca II absorption was injected.The top plot shows a negative artificial spectrum, and the bottom plot shows a positive one. Both of these come from the same DR12 quasar spectra, with the only difference being the injected lines. Note that the top plot shows the spectra after 10 pixels around each line location have been replaced with white noise. This negative spectra is among the 70\% of total spectra that have neither line injected. The other 30\% would either have only the first line injected, or the second line injected.}
 \label{fig:ca_inj_demo}
\end{figure}

Due to the need for large sets of training data and the scarcity of real Ca II absorption lines that have been already discovered, generating artificial training data that closely mimics genuine conditions was essential to this study. To accomplish this, we built upon techniques established in prior research, such as \cite{2019MNRAS.487..801Z} and \cite{2022MNRAS.517.4902X}.

The process begins with randomly generating absorption redshifts ($z_{\text{abs}}$) to determine the location of injection. Then equivalent widths (EWs) and full width at half maximum (FWHM) of the Ca II lines are modeled using log-normal distributions measured from the catalog given by \cite{2022MNRAS.517.4902X}. The EW of the first Ca II line at 3934 \AA\ is randomly drawn from this distribution, while the EW of the second line is calculated based on the distribution of the measured ratio between the first and second lines' EWs in the data. The comparisons between the injected EW and measured EW are shown in Figure~\ref{fig:EW_distributions}. 

To ensure that the injected lines are not superimposed on existing spectral features, such as other absorption or emission lines, the flux in a 10-pixel window centered on each line is replaced with the continuum fit in this window. Then, noise generated from a gaussian distribution scaled to the median of local flux error values is added. 

Since the $z_{\text{abs}}$ value reported from Mg II absorption may not always be accurate, often due to contamination of the Mg II lines, we introduce random offsets to the $z_{\text{abs}}$ values associated with the Ca II lines. These offsets are sampled from a gaussian distribution with a standard deviation of 0.00015 in redshift, a parameter derived from the results of \cite{2022MNRAS.517.4902X}. Additionally, a rest-frame offset of $0.15\,\text{\AA}$ is applied uniformly to each Ca II line separately to further account for any systematic errors.

Finally, each training sample is fitted by a gaussian and the SNR of both lines are measured. If the SNR of the first line is less than $3.0$ or the SNR of the second line is less than $2.5$, the sample is discarded. This ensures that no bad samples with small lines are generated.

For negative training samples in the CNN, areas corresponding to the Ca II lines are also replaced with white noise in the same fashion as the positive samples. This bypasses the risk of having another absorption line coincidentally appear in that region. Furthermore, to ensure robust classification, 30\% of these samples include exactly one randomly injected line, with each line being equally likely to appear. This trains the CNN to differentiate effectively between genuine Ca II absorption lines and noise or incomplete line features with only one absorption-like feature, thereby enhancing the accuracy of the model. Examples of these positive and negative samples are shown in Figure~\ref{fig:ca_inj_demo}.

Following line injection, the spectra is cropped to a length of 81 pixels centered at the mean wavelength of the two lines. Local continuum fitting is then applied to this section. Noise normalization is then performed; here, the error is measured using \texttt{astropy}'s sigma clipped stats, explicitly excluding the ten pixels surrounding each injected line to avoid bias from the absorption features. After noise normalization, we divide the spectral values by 20, and then add 0.5. This puts the flux values in the range 0 to 1, with the continuum at 0.5. This is done because many activation functions operate in the range 0 to 1, and all the data that is fed into the CNN will be processed in the same way. Real spectra are processed in a similar fashion, with the same cropping, continuum fitting, and normalization steps applied.

\subsection{Artificial Fe II Data Set}

\begin{figure}
 \includegraphics[width=\columnwidth]{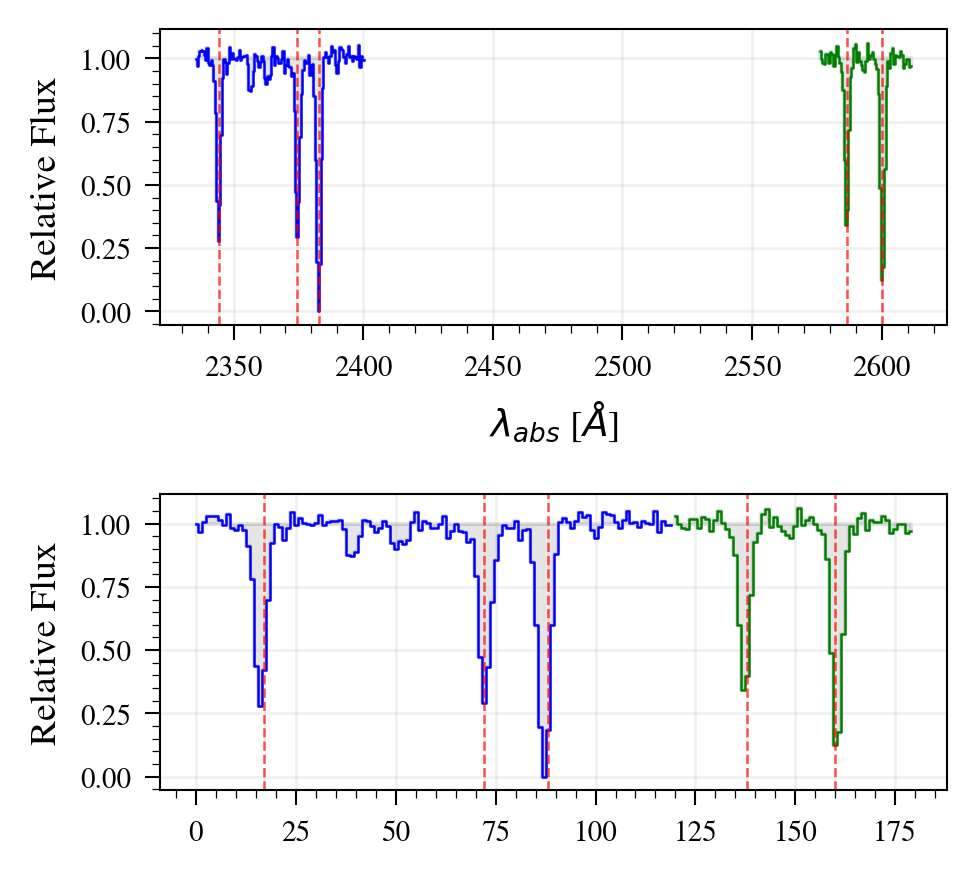}
 \caption{A demonstration of the process of cropping and concatenating two windows such that the five Fe II lines are joined and unnecessary data is removed to create a faked quintuplet for CNN training and Fe II line detection. }
 \label{fig:feii_window_demo}
\end{figure}

\begin{figure*}
 \includegraphics[width=\textwidth]{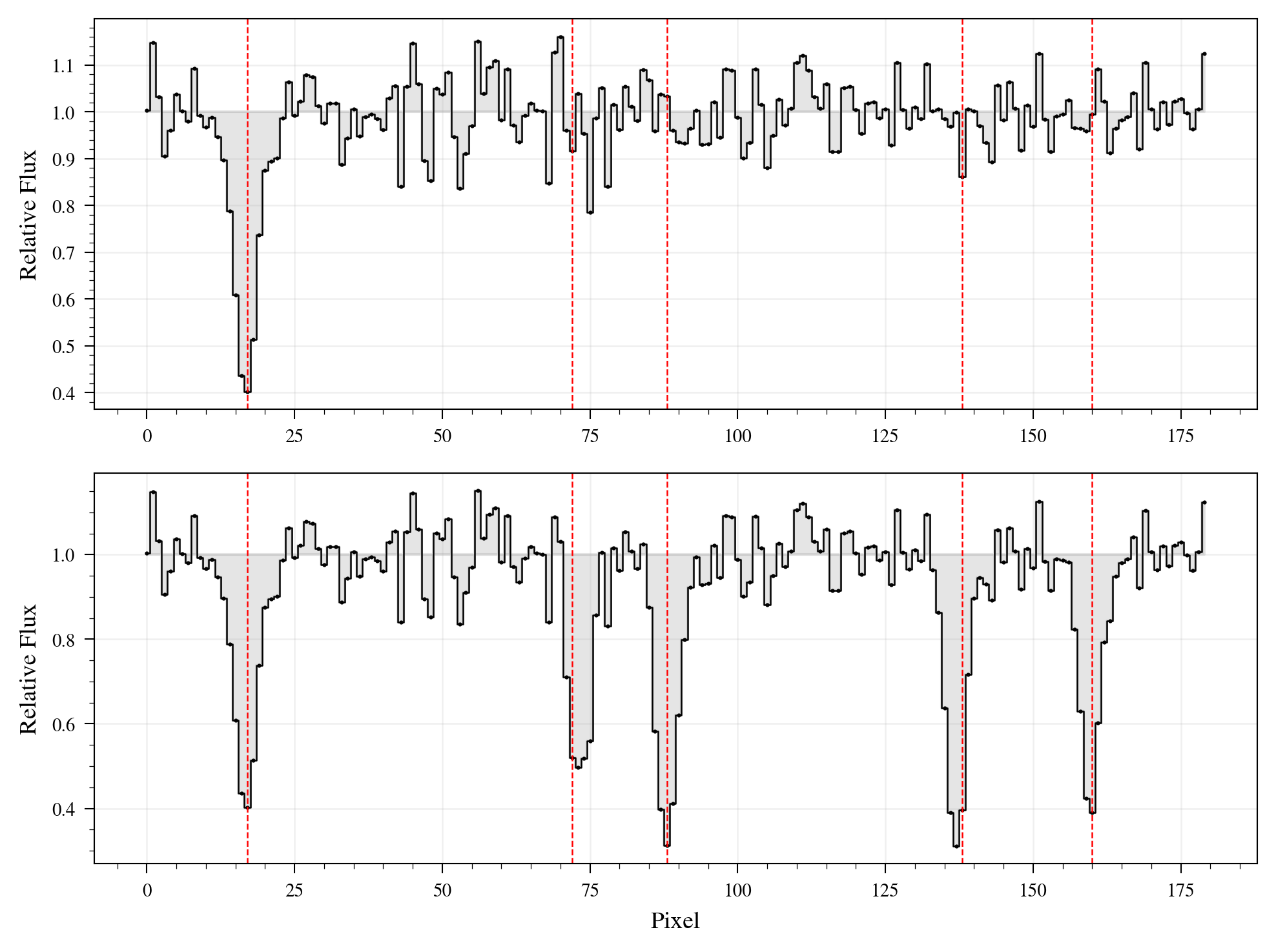}
 \caption{Samples of artificial spectra that are generated for Fe II. The red dotted lines indicate the locations at which Fe II absorption was injected. The top plot shows the negative artificial spectrum, while the bottom plot shows the positive one. Both of these come from the same DR12 quasar spectra, with the only difference being pixels around the injected lines. This negative spectrum is among the 30\% of total spectra that have a subset of the lines injected. Specifically, this one had exactly one line injected, namely the $\lambda$ 2344 Fe II line.}
 \label{fig:fe_inj_demo}
\end{figure*}

The injection process for Fe II lines mirrors that of Ca II, utilizing data from the \cite{2022MNRAS.517.4902X} catalog to model the EW and FWHM of these lines. The strongest Fe II line at 2382 \AA, chosen based on its high oscillator strength value from \cite{2014MNRAS.438..388M}, serves as the primary reference for injection. Its EW is randomly drawn from its measured lognormal distribution. Subsequently, the EWs of the remaining four Fe II lines at wavelengths 2344, 2374, 2586, and 2600 \AA\ are determined by multiplying the primary line's EW by ratios randomly sampled from their respective lognormal distributions, derived from measured EW ratios relative to the 2382 \AA\ line. This approach accurately captures the relative strengths and inherent variability of Fe II lines observed in real spectra.

To capture the necessary spectral features efficiently, we use a cropping and concatenating technique inspired by the "faked doublet method" from \cite{2024MNRAS.531..387G}. The input to the CNN consists of two segments concatenated together. The first segment is centered around the first three lines ($\lambda \lambda$ 2344, 2374, 2382), with a window length of 120 pixels. The second segment surrounds the doublet at $\lambda \lambda$ 2586 and 2600 with a window length of 60 pixels. Figure~\ref{fig:feii_window_demo} illustrates how this process is done. Each window then undergoes individual continuum fitting, noise normalization, and special CNN normalization using the same methods and packages as those applied to the Ca II data. Again, the real spectra are cropped, combined, and normalized in a similar manner.

This segmentation allows for focused analysis of the regions of greatest interest without sacrificing relevant spectral information. The real spectra are also processed similarly, ensuring that the CNN consistently handles both synthetic and real data.

Similar to the Ca II training samples, for the false examples in the Fe II data set, we employ a partial injection strategy where 30\% of the data may contain up to three randomly selected Fe II lines. This setup is intended to challenge the neural network's ability to accurately identify Fe II lines, thereby training it to label spectra as true only if they have all five lines. Examples of these positive and negative samples are shown in Figure~\ref{fig:fe_inj_demo}.

\subsection{Model Structure}

\begin{figure*}
 \includegraphics[width=\textwidth]{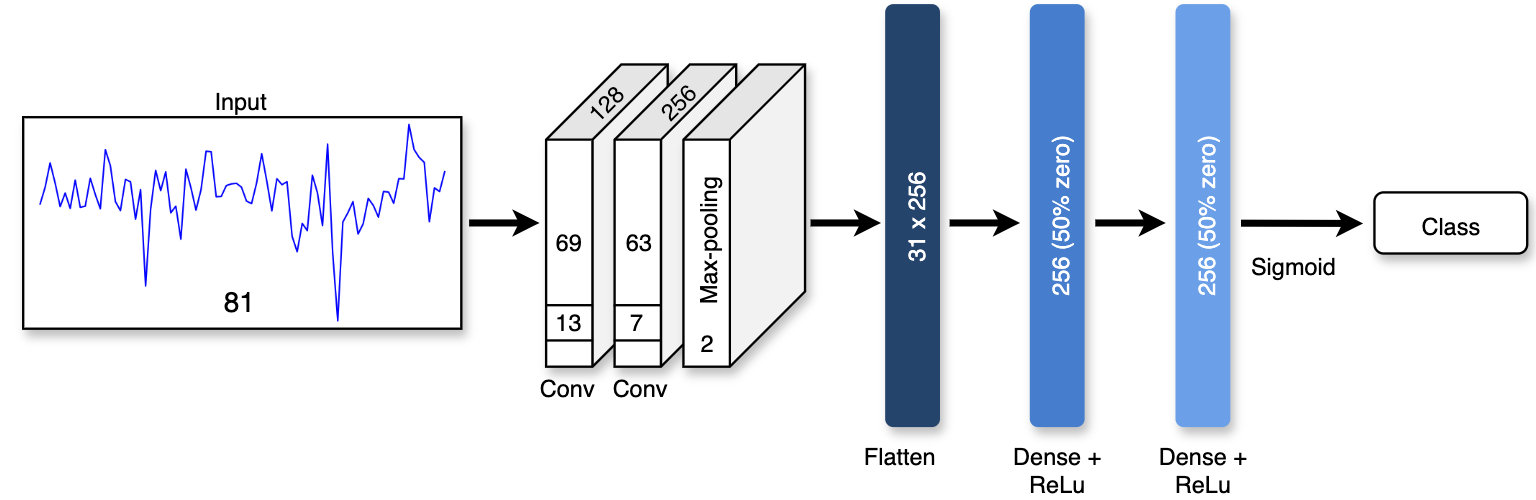}
 \caption{Schematic of our CNN architecture, which consists of 2 convolutional layers followed by two dense layers, each with 50\% dropout which was only active during training.}
 \label{fig:cnn_architecture}
\end{figure*}

Both our neural networks are designed to identify features that are consistently at specific locations within a 1D window. Therefore, they consist of similar components. After considerable testing, we settled on the best performing structures for the neural networks.

Our Ca II CNN model, shown in Figure~\ref{fig:cnn_architecture}, is structured to process 1D data inputs of size 81, starting with two convolutional layers that extract important features from the input data. The first layer applies 128 filters with a kernel size of 13, which works to detect larger patterns in the input, followed by a second convolutional layer with 256 filters and a smaller kernel size of 7 to capture finer details. Both convolutional layers use a linear activation function. After the convolutional layers, a max pooling layer with a pool size of 2 reduces the spatial dimensions of the output, downsampling the feature maps. The output is then flattened to ensure that the data can be properly fed into the subsequent dense layers. We have two dense layers in total, each with 256 units and Rectified Linear Unit (ReLU) activation. A 50\% dropout rate is applied after each dense layer to introduce regularization and reduce overfitting. Finally, a single neuron in the output layer, using a sigmoid activation function, provides a scalar output ranging from 0 to 1.

Our Fe II CNN model is of similar structure, the only difference being the input is 1D data of size 180, and the convolutional layers have kernel sizes 3 and 7 respectively.

\subsection{Model Training}

\begin{figure}
 \includegraphics[width=\columnwidth]{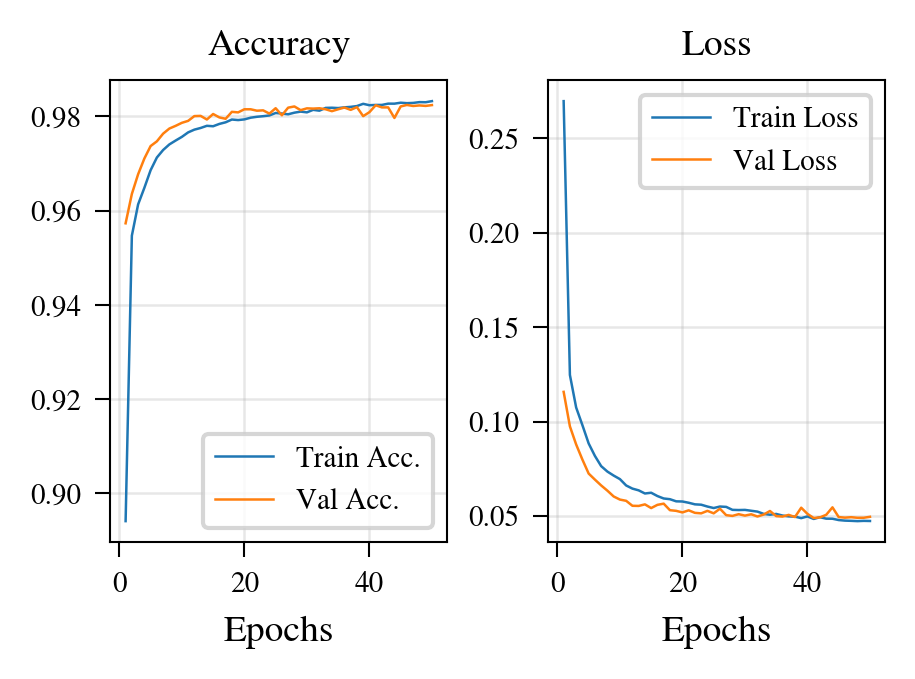}
 \caption{Ca II identification model accuracy and loss throughout the epochs.}
 \label{fig:ca_train}
\end{figure}

\begin{figure}
 \includegraphics[width=\columnwidth]{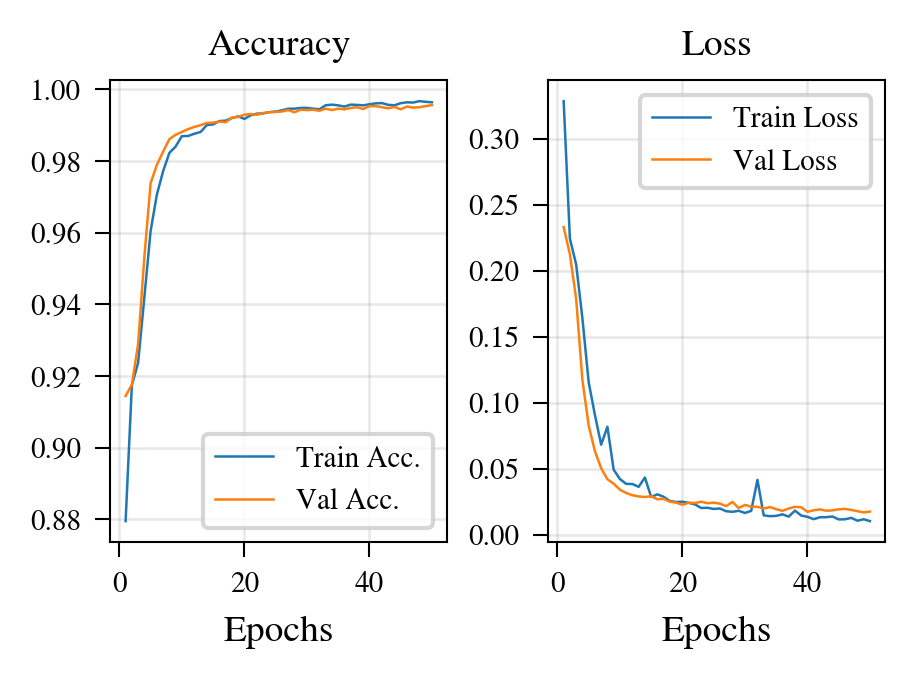}
 \caption{Fe II identification model accuracy and loss throughout the epochs.}
 \label{fig:fe_train}
\end{figure}

\begin{figure}
 \includegraphics[width=\columnwidth]{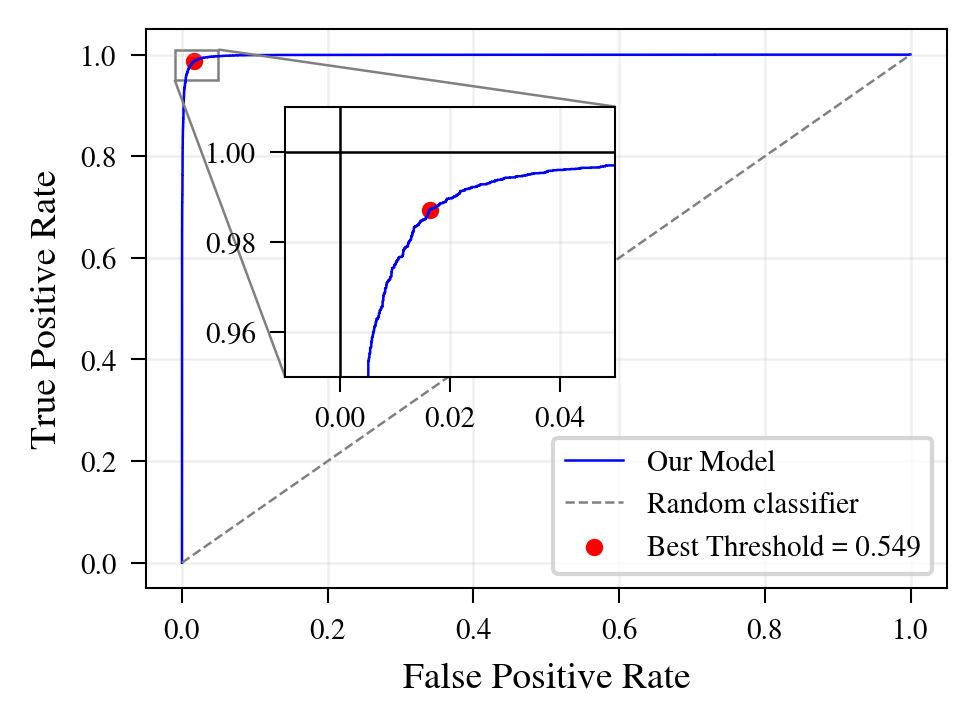}
 \caption{The ROC Curve of the Ca II model, which shows its sensitivity and specificity for various thresholds.}
 \label{fig:ca_roc}
\end{figure}

\begin{figure}
 \includegraphics[width=\columnwidth]{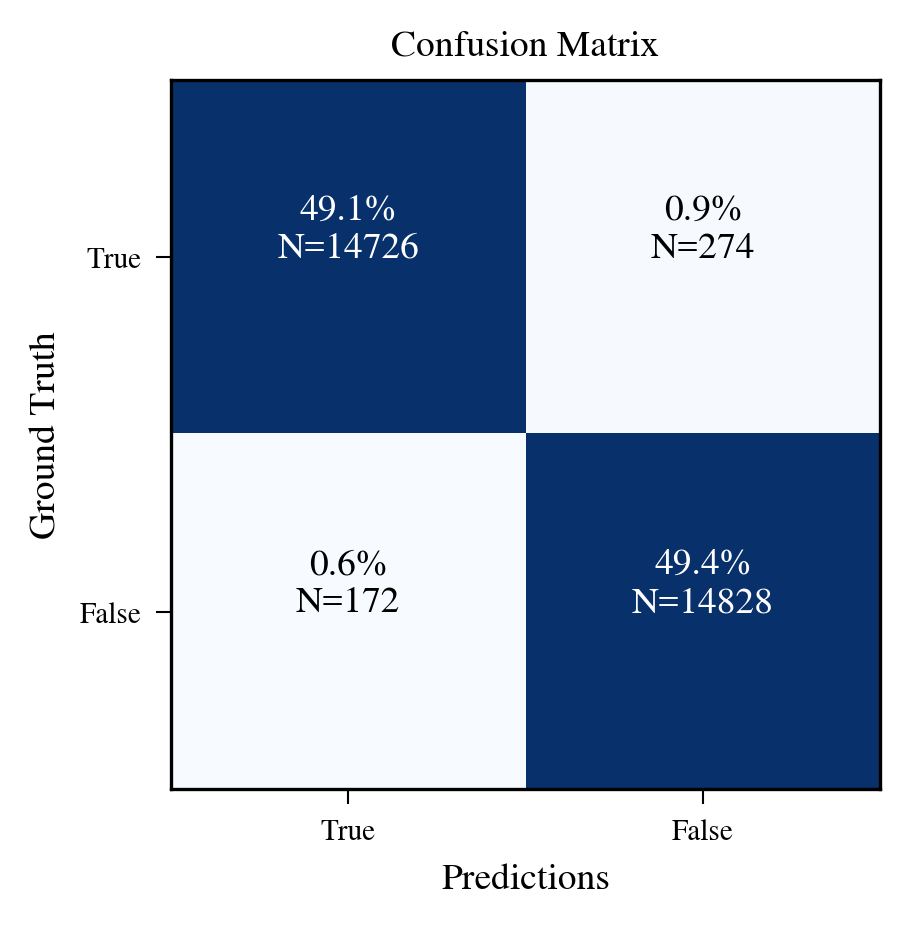}
 \caption{The confusion matrix of our Ca II model’s predictions of a test set of 30,000 synthetic injections. The model had false positive and false negatives rates of 0.9\% and 0.6\% respectively.}
 \label{fig:ca_conf}
\end{figure}

\begin{figure}
 \includegraphics[width=\columnwidth]{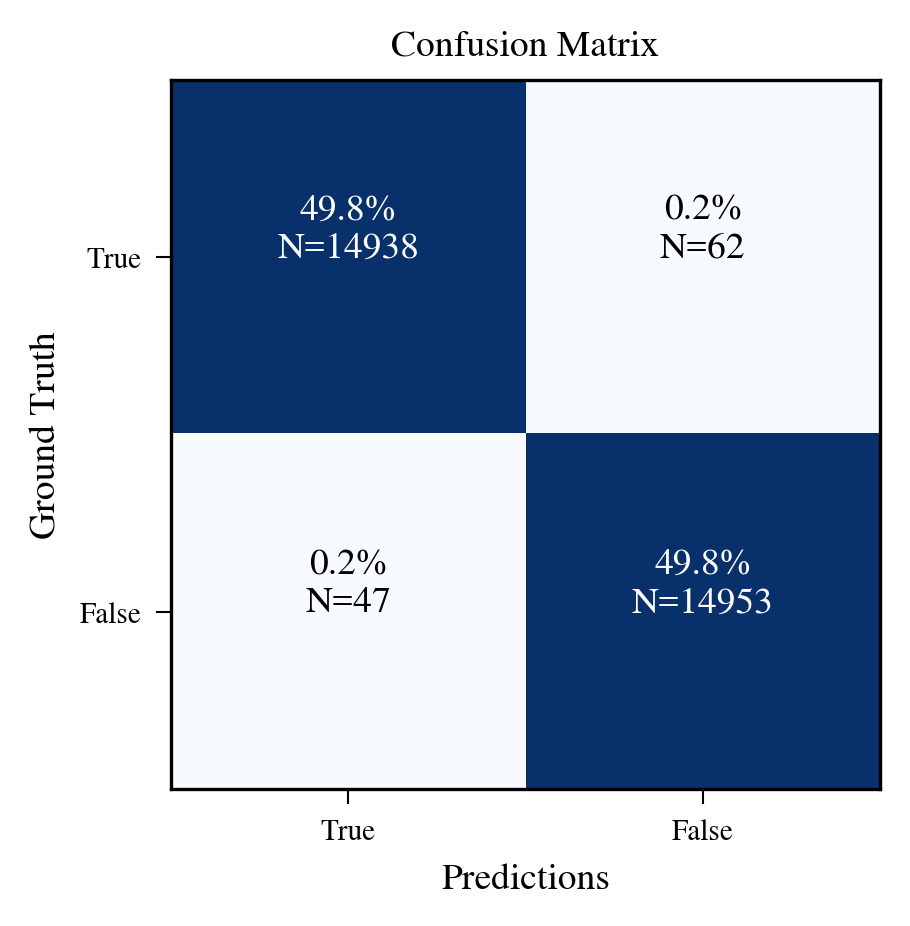}
 \caption{The confusion matrix of our Fe II model’s predictions of a test set of 30,000 synthetic injections. The model had false positive and false negatives rates of 0.2\% and 0.2\% respectively.}
 \label{fig:fe_conf}
\end{figure}

In total, we had 200,000 positive samples and 200,000 negative samples for Ca II absorption and the same number for Fe II absorption, tallying up 400,000 samples for each CNN. Of these, 300,000 samples were allocated for training, 70,000 for validation, and 30,000 for testing, ensuring an equal distribution of positive and negative samples across all sets. The Adam optimizer was used for both \citep{kingma2017adammethodstochasticoptimization}. The loss function for both neural networks is the standard binary cross-entropy loss.

Both models were trained using an NVIDIA RTX 3090 Ti graphics card, running for 50 epochs on the Ca II model and 40 epochs on the Fe II model. During each epoch, the training samples were processed in batches of 32, with a buffer size of 320, allowing for efficient memory management and fast training. The accuracy and loss metrics for the Ca II and Fe II models across the epochs are illustrated in Figure~\ref{fig:ca_train} and Figure~\ref{fig:fe_train} respectively. After the training phase, the models were evaluated on the test set consisting of 30,000 samples. The receiver operating characteristic (ROC) curve of our CA II model is shown in Figure~\ref{fig:ca_roc}. The optimal threshold from this was measured to be $0.55$. For simplicity, we set our threshold to $0.5$, as there was almost no difference. The results for each model are summarized in the confusion matrix shown in the Figure~\ref{fig:ca_conf} and Figure~\ref{fig:fe_conf}. The neural network trained on Ca II data achieved an accuracy of 98.4\%, while the Fe II model reached an accuracy of 99.8\%. These results highlight the exceptional performance of the models, demonstrating their ability to accurately classify spectral data, even in the presence of complex and unpredictable spectral features.

\section{Results}
\label{sec:Results}

\subsection{Results on Existing Catalog}

\begin{figure}
 \includegraphics[width=\columnwidth]{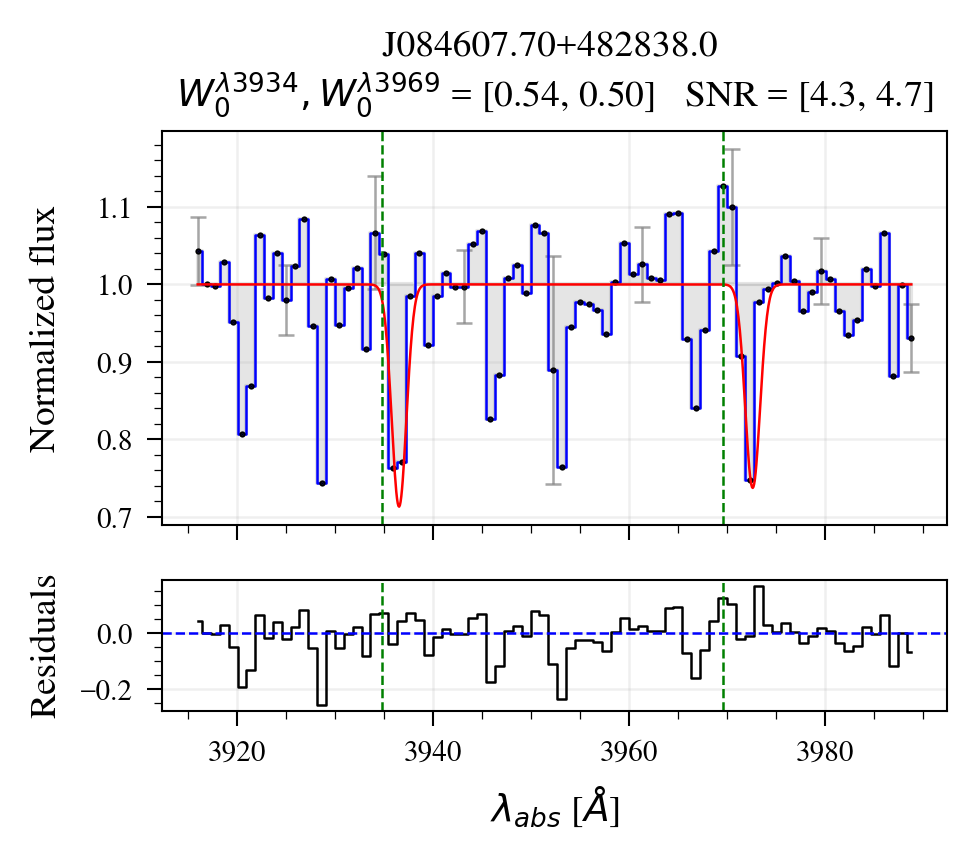}
 \caption{The one spectrum in our positive test set that was missed by the Ca II CNN. The Ca II $\lambda\lambda$ 3934, 3969 lines appear weak and slightly offset, with significant noise across the region. The model assigned a confidence score of 0.46—just below the detection threshold of 0.5—resulting in a false negative.}
 \label{fig:ca_bad}
\end{figure}

Among the 305 spectra in our positive test set from \cite{2022MNRAS.517.4902X}, our Ca II neural network successfully identified 304 as containing Ca II absorption features. Conversely, among the 305 spectra in our randomly generated negative test set, the CNN labeled all entries as devoid of Ca II absorption lines. This corresponds to an overall accuracy of 99.8\%.

The one spectrum our model failed to detect is shown in Figure~\ref{fig:ca_bad}. It appears to be a marginal detection, with significantly offset lines and a high level of noise. The model assigned it a confidence score of 0.46, just below our detection threshold of 0.5.

\subsection{Results on DR16}

\begin{table*}
\centering
\caption{Randomly selected 10 of the 1121 new Ca II detections in the DR16 Mg II catalog; the rest are listed as supplementary information.}
\label{tab:caii_dr16_ex}
\begin{tabular}{lccccccccc}
\hline
SDSS Name & RA & Dec. & MJD & Plate & Fiber & $z_{\text{qso}}$ & $z_{\text{abs}}$ & $W_0^{\lambda 3934}$(\AA) & $W_0^{\lambda 3969}$(\AA)\\
\hline
J015652.62\texttt{+}274551.3 & 29.219 & 27.764 & 58462 & 10000 & 661 & 1.051 & 0.500 & $0.8 \pm 0.1$ & $0.5 \pm 0.2$ \\
J093543.38\texttt{+}201403.2 & 143.931 & 20.234 & 57805 & 9560 & 142 & 1.011 & 0.450 & $1.2 \pm 0.2$ & $0.5 \pm 0.2$ \\
J105644.97\texttt{+}362351.0 & 164.187 & 36.397 & 58220 & 10269 & 764 & 1.867 & 0.756 & $0.3 \pm 0.1$ & $0.4 \pm 0.2$ \\
J114902.70\texttt{+}144328.1 & 177.261 & 14.724 & 56009 & 5381 & 874 & 2.200 & 1.103 & $0.16 \pm 0.06$ & $0.18 \pm 0.08$ \\
J120155.31\texttt{+}364442.9 & 180.480 & 36.745 & 57428 & 8864 & 274 & 1.853 & 1.384 & $1.9 \pm 0.8$ & $1.9 \pm 0.9$ \\
J134517.08\texttt{+}400054.8 & 206.321 & 40.015 & 57429 & 8847 & 646 & 0.938 & 0.477 & $1.1 \pm 0.3$ & $0.7 \pm 0.2$ \\
J161338.14\texttt{+}443933.0 & 243.409 & 44.659 & 57900 & 8525 & 678 & 1.793 & 0.676 & $0.8 \pm 0.2$ & $0.3 \pm 0.2$ \\
J215810.11\texttt{+}055933.6 & 329.542 & 5.993 & 58431 & 11337 & 652 & 2.625 & 1.253 & $0.4 \pm 0.2$ & $0.3 \pm 0.1$ \\
J225034.26\texttt{+}333157.7 & 342.643 & 33.533 & 56565 & 7141 & 838 & 2.203 & 1.135 & $1.8 \pm 0.6$ & $1.7 \pm 0.3$ \\
J235301.94\texttt{-}001157.5 & 358.258 & -0.199 & 57666 & 9159 & 606 & 1.252 & 0.372 & $0.9 \pm 0.2$ & $0.3 \pm 0.1$ \\
\hline
\end{tabular}
\end{table*}

\begin{figure*}
 \includegraphics[width=\textwidth]{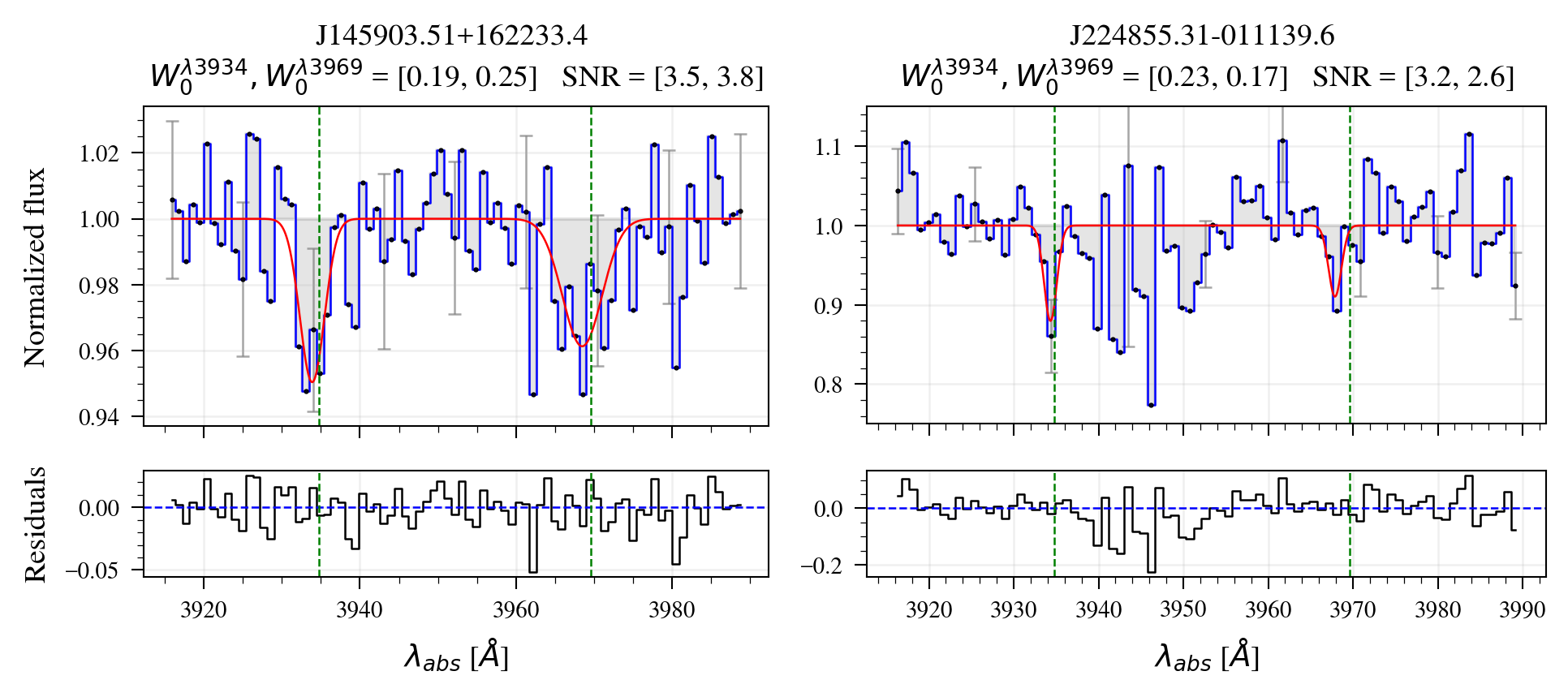}
 \caption{Examples of two spectra that were removed in our manual selection process. The spectra shown in the left had significant blending in both lines. The spectra shown in the right plot had fairly weak lines that were also offset.}
 \label{fig:dr16_bad_ex}
\end{figure*}

\begin{figure}
 \includegraphics[width=\columnwidth]{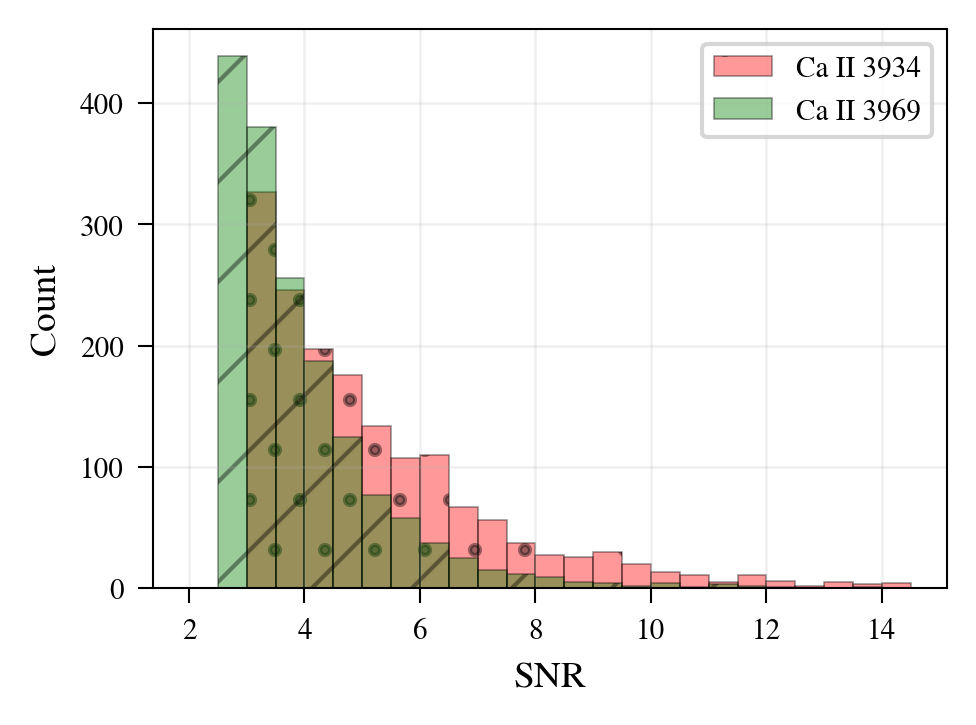}
 \caption{The distribution of SNR for both lines across our Ca II absorption catalog.}
 \label{fig:snr_dist}
\end{figure}

\begin{figure*}
 \includegraphics[width=\textwidth]{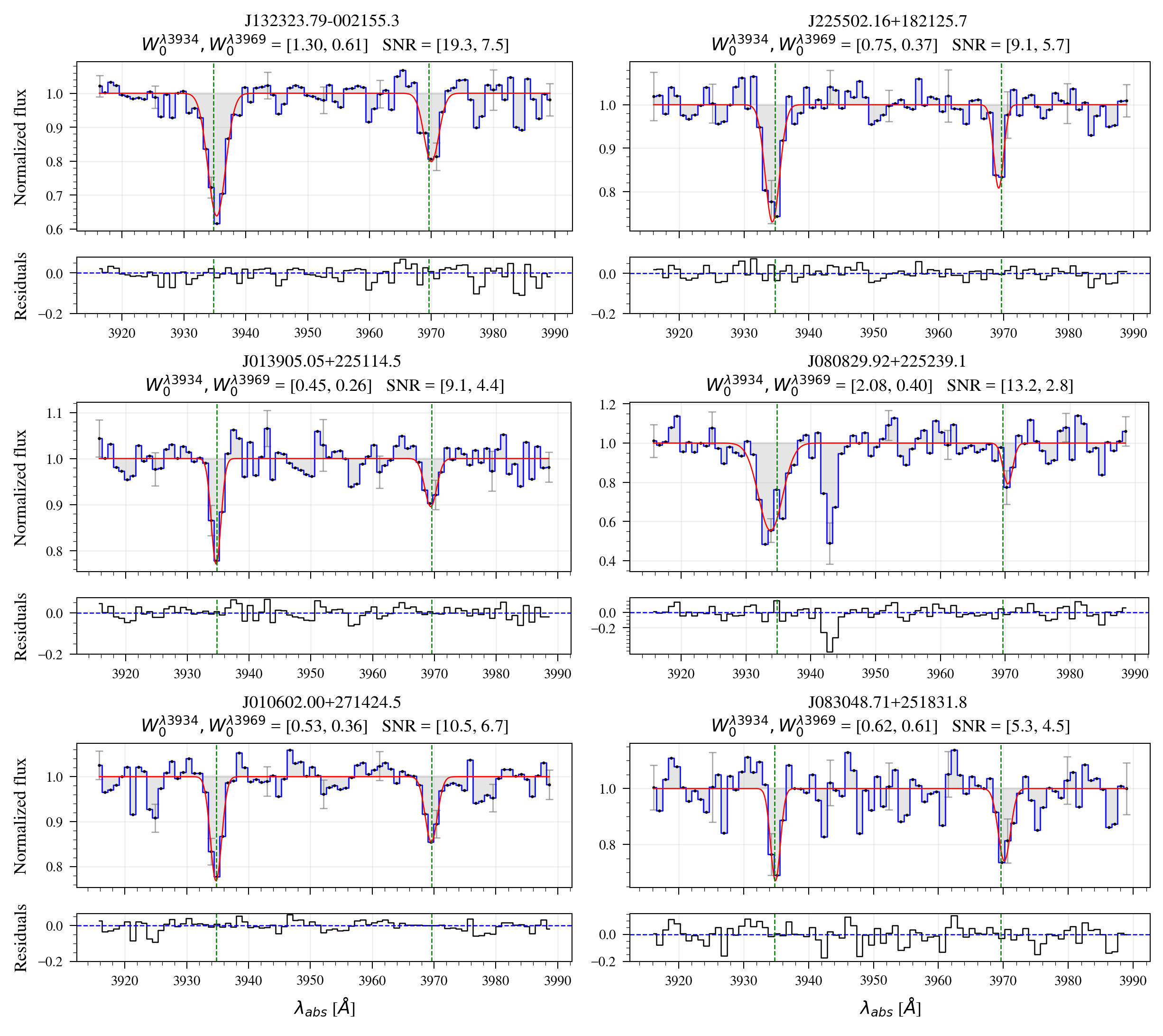}
 \caption{Six examples among the 1121 new Ca II absorption detections by our pipeline.}
 \label{fig:dr16_ex_detections}
\end{figure*}

\begin{figure*}
 \includegraphics[width=\textwidth]{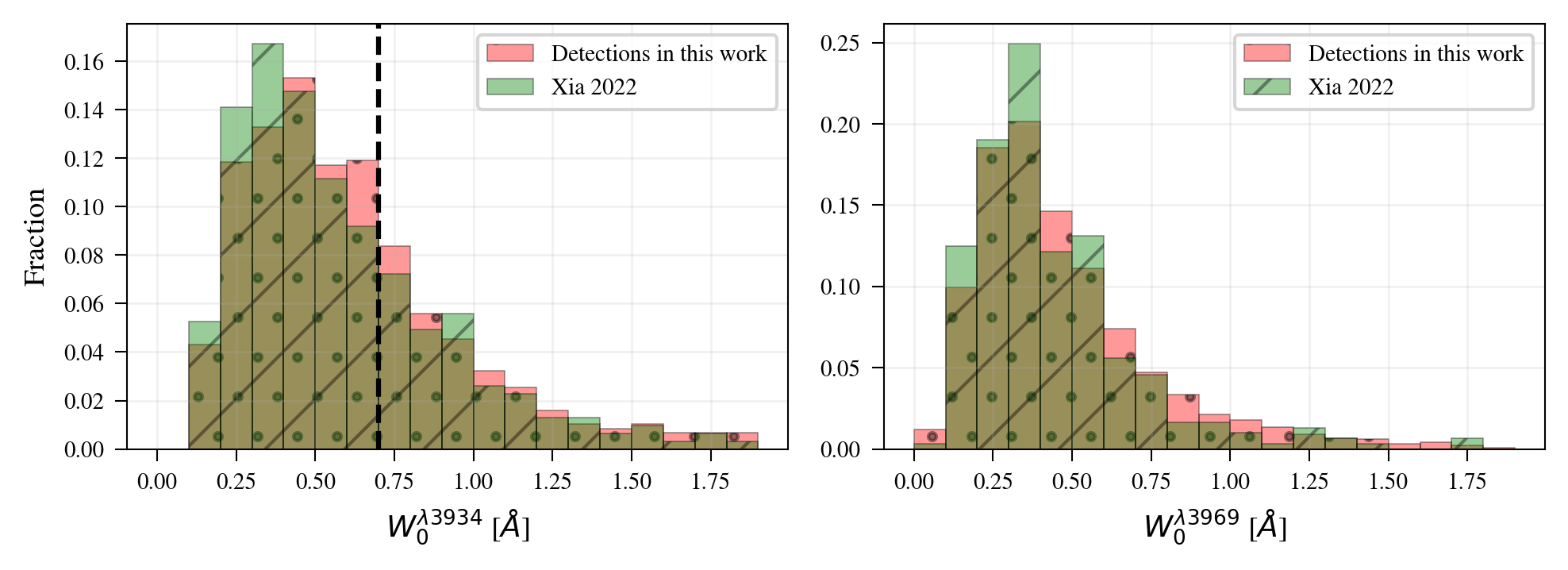}
 \caption{The distribution of the rest-frame equivalent widths of $\lambda \lambda$ 3934 and 3969 from our DR16 absorption detections compared to that of \citet{2022MNRAS.517.4902X}. Both catalogs resemble a similar trend, with the majority of the detections being weak ($W_0^{\lambda 3934} < 0.7$). On the left plot, the black dotted line represents the split at $W_0^{\lambda 3934} = 0.7$ between strong and weak detections. Our new catalog contains 520 strong detections and 1126 weak detections.} 

 \label{fig:ew_cmp_dr16_iona}
\end{figure*}

\begin{figure}
 \includegraphics[width=\columnwidth]{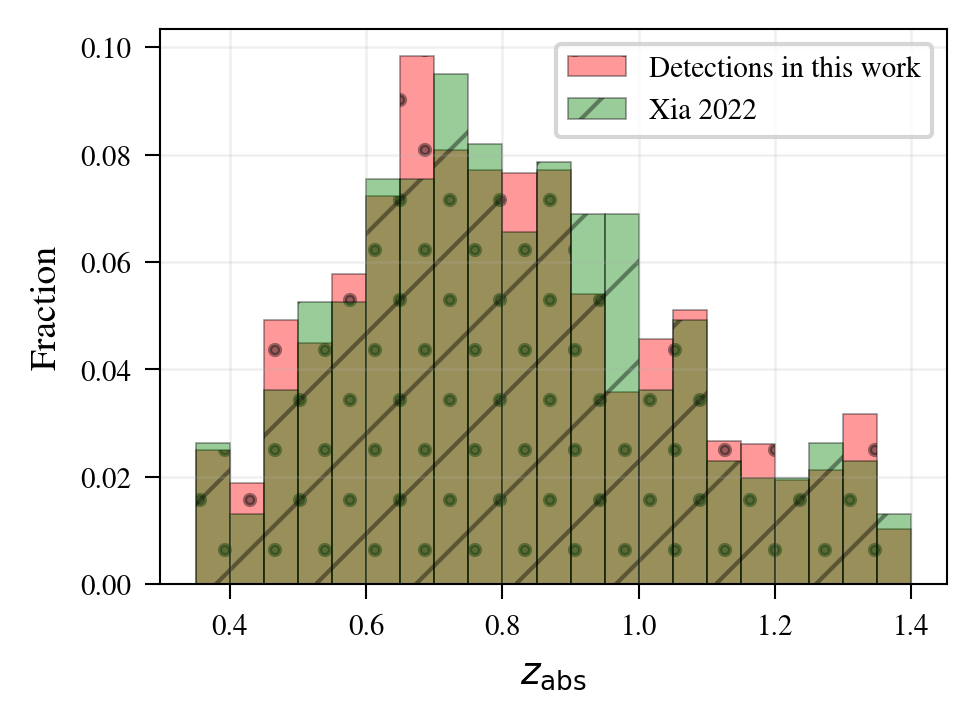}
 \caption{The distribution of the absorption redshifts of our DR16 detections compared to that of \citet{2022MNRAS.517.4902X}. }
 \label{fig:zabs_dist_cmp}
\end{figure}

\begin{figure}
 \includegraphics[width=\columnwidth]{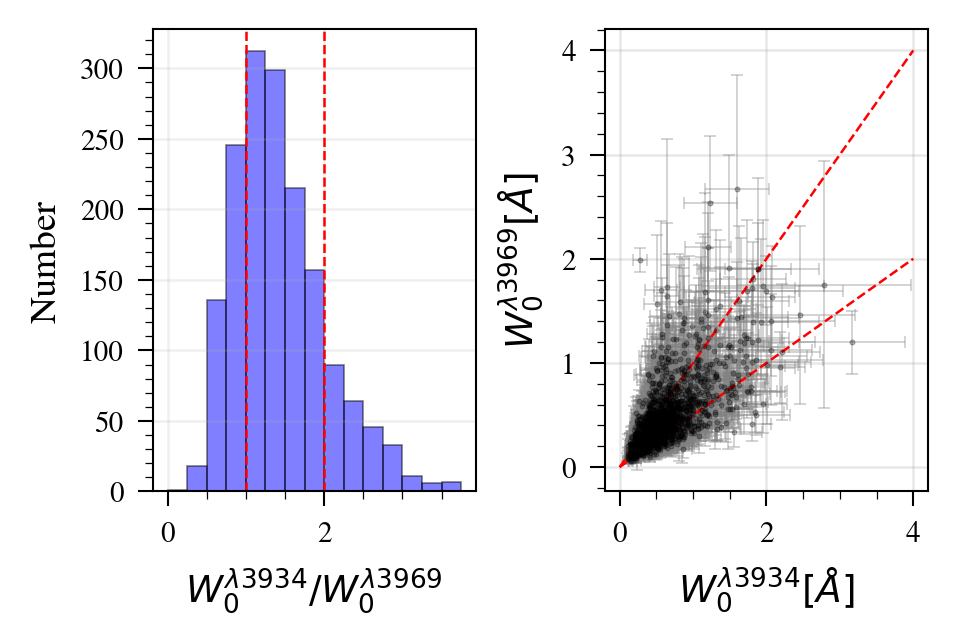}
 \caption{The left plot shows the distribution of $W_0^{\lambda3934} / W_0^{\lambda 3969}$ across our catalog, and the right plot shows $W_0^{\lambda 3969}$ plotted against $W_0^{\lambda3934}$. The red lines in both plots represent $W_0^{\lambda3934} / W_0^{\lambda 3969} =$ 1 and 2, which are the limits for fully saturated and unsaturated absorption systems.}
 \label{fig:w_ratio}
\end{figure}

Utilizing our dual CNN approach, we analyzed a substantial data set of 108,783 quasars from the Mg II absorption catalog provided by \cite{2021MNRAS.504...65A}. This subset of their full catalog, characterized by redshifts in the range $0.36 < z_{\text{abs}} < 1.4$, was methodically processed to identify and verify Ca II absorption features. The procedure comprised several meticulously designed steps, detailed as follows:

\begin{enumerate}
    \item \textbf{Initial Ca II identification:} Our primary CNN, trained as a binary classifier to identify Ca II absorption lines, was employed to assign a score between 0 and 1 for each quasar spectrum. We applied an initial filter to retain spectra with scores exceeding 0.5, suggesting the probable presence of Ca II lines. This filter reduced the sample size to 6,667 candidates deemed worthy of further analysis.
    
    \item \textbf{Fe II cross-identification:} Simultaneously, a secondary CNN specialized in detecting Fe II absorption lines evaluated the same quasar spectra, providing a complementary verification method. However, spectra were only evaluated by the Fe II model if they satisfied the following criteria: (a) The Mg II absorber redshift exceeded 0.7063, ensuring the Fe II $\lambda2344$ line appeared at wavelengths longer than 4000 Å, thus avoiding the noisy data regions at shorter wavelengths; and (b) All five Fe II absorption lines in question were located outside the Ly$\alpha$ forest. 
    This selection step reduced the candidate list to 5,390 spectra, demonstrating the effectiveness of using Fe II to minimize false positives at higher redshifts and enhance detection reliability.
    
    \item \textbf{Spectral line fitting and SNR calculation:} For the remaining candidates, we conducted rigorous line fitting using a 1D Gaussian model for each of the Ca II absorption lines. The signal-to-noise ratio for each line was calculated using the following method, which provides a robust measure of line visibility against the background noise:

    \begin{equation}
        \text{SNR} = \sqrt{\sum \left(\frac{1 - F_{\text{model}}}{\sigma_{\text{error}}}\right)^2}
    \end{equation}
    
    $F_{\text{model}}$ represents the Gaussian fit of the spectral line after continuum fitting is applied, and $\sigma_{\text{error}}$ is the array of normalized flux errors provided by SDSS. We adopted SNR thresholds of 3.0 for $\lambda 3934$ and 2.5 for $\lambda 3969$, based on the criteria established by \cite{2022MNRAS.517.4902X}. This step reduced the number of detections to 1,924.
    
    \item \textbf{Manual Selection: } After the automated identification and initial screening, we conducted a detailed manual review of the remaining Ca II absorption candidates to ensure the quality of the detections. This involved carefully inspecting each spectrum to identify and remove those with significant contamination from nearby spectral features, such as overlapping absorption lines or emission artifacts that could obscure the Ca II doublet. We also checked for any wavelength misalignment between the two lines of the Ca II doublet, removing spectra where one line was significantly offset from the expected position. Spectra with unusually weak or poorly defined absorption features that could not be reliably distinguished from noise were also excluded. Figure~\ref{fig:dr16_bad_ex} shows 2 such spectra that were removed. This thorough review removed 233 spectra and reduced the number of Ca II absorption systems in the final catalog to 1,646 detections. Figure~\ref{fig:snr_dist} shows the distributions of the SNRs of each line in the Ca II doublet.
\end{enumerate}

The final catalog of 1,646 Ca II absorption detections represents a significant expansion of the existing body of known Ca II absorbers when compared to the previous catalogs produced. Among these 1,646 absorbers, 525 were already present in previous research \citep{2014MNRAS.444.1747S, 2022MNRAS.517.4902X, 2023MNRAS.518.5590F, 2025ApJS..276...37L}. By cross-referencing all 5 catalogs and removing duplicates, we established a final data set of 2,143 unique Ca II absorbers. This consolidated catalog represents the most extensive compilation of Ca II absorbers to date. However, to represent a homogeneous data set of Ca II absorbers among Mg II absorbers, only the 1,646 of our absorbers will be used for the statistical analyses that follow. Figure ~\ref{fig:dr16_ex_detections} contains 6 new absorbers that our pipeline identified. 

A comparison of the rest-frame EW distributions of the Ca II $\lambda \lambda$ 3934 and 3969 lines between our catalog and that of \cite{2022MNRAS.517.4902X} is shown in Figure~\ref{fig:ew_cmp_dr16_iona}. Both catalogs follow a similar trend, with the majority of detections classified as weak absorbers. This predominance of weak absorbers is consistent with previous studies, suggesting that Ca II absorption systems with higher equivalent widths are rare. In terms of absorption redshift, the distribution of our detections and those of \cite{2022MNRAS.517.4902X} is shown in Figure~\ref{fig:zabs_dist_cmp}.

The rest-frame EW is calculated with $\sqrt{2\pi}\cdot A\cdot\sigma$ where $A$ is the amplitude of the Gaussian fit and $\sigma$ is the standard deviation of the fit. Figure ~\ref{fig:w_ratio} shows the distribution of DR ($W_0^{\lambda3934} / W_0^{\lambda 3969}$) in our catalog. EW width error, following the methods of \cite{2023MNRAS.518.5590F}, is given by $\text{FWHM}\cdot\sigma_\text{rms}$ where $\text{FWHM}$ is the measured rest-frame FWHM from the Gaussian, and $\sigma_\text{rms}$ is the mean flux noise in a 2.5 \AA\ window around the line center given by SDSS.

Table ~\ref{tab:caii_dr16_ex} contains 10 random absorbers from our total 1121 new discoveries. The rest are given as supplementary material. 

\subsection{Equivalent Width Completeness Analysis}
\label{sec:Equivalent_Width_Analysis}

\begin{figure}
 \includegraphics[width=\columnwidth]{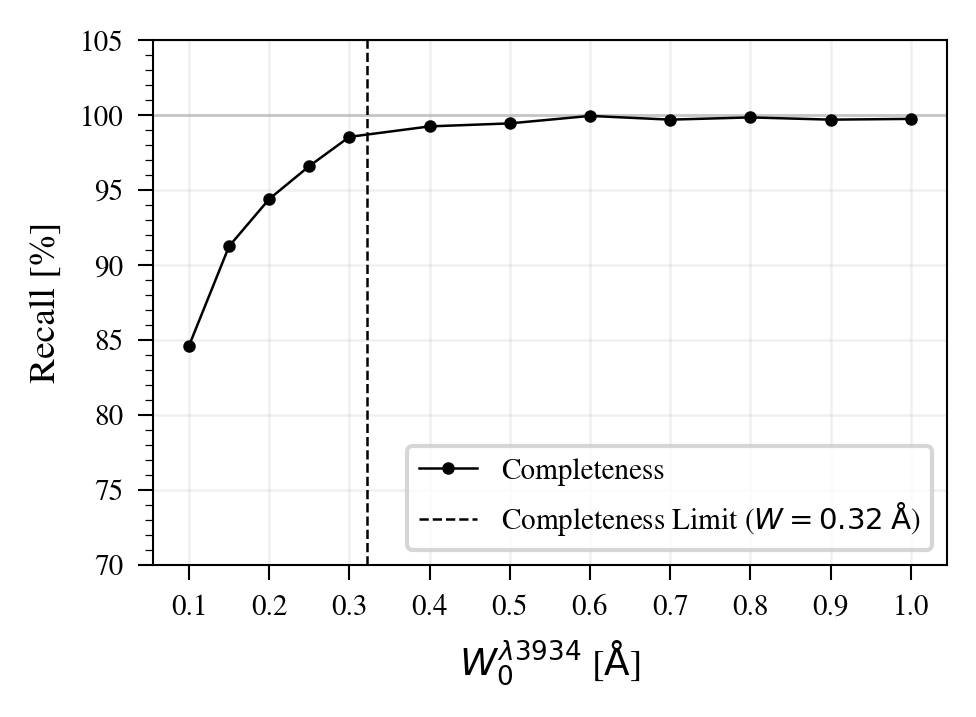}
 \caption{The recall performance of our Ca II CNN across varying rest-frame equivalent widths for the $\lambda 3934$ line. Samples are generated by injecting synthetic absorption lines to SDSS DR12 spectra, using the same algorithm used to generate training samples for the CNN. Using the 99\% recall threshold, we calculate the $W$ limit of our Ca II CNN to be 0.32\AA.}
 \label{fig:w_completeness}
\end{figure}

We determine the completeness and the $W$ limit of our Ca II CNN with a method from \cite{2015A&A...580A...8L}. The injected EW of the $\lambda 3934$ line ranges from $0.1$ to $1$ \AA, with $0.1$ \AA\ increments. The EW of $\lambda 3969$ is modeled with the distribution of the ratio between the EW of the two lines, measured from \cite{2022MNRAS.517.4902X}. The line injection algorithm in this process is the same one used to generate our CNN training samples. In addition, we included 2 additional bins at $W = 0.15$ and $0.25$ to better estimate the completeness at lower equivalent widths. In each of these 12 bins, 2000 true samples are created and tested. 

Similar to \cite{2024MNRAS.531..387G}, our completeness limit was defined as the lowest EW at which the recall is at least 99\%. However, unlike \cite{2024MNRAS.531..387G}, our CNN score threshold was set to $0.5$ rather than $0.94$, matching the value used in our DR16 search. Figure~\ref{fig:w_completeness} illustrates the recall at each bin, as well as the EW limit, measured to be $0.32$ \AA.

\subsection{Composite Spectra}
\label{sec:Composite_Spectra}

\begin{figure}
 \includegraphics[width=\columnwidth]{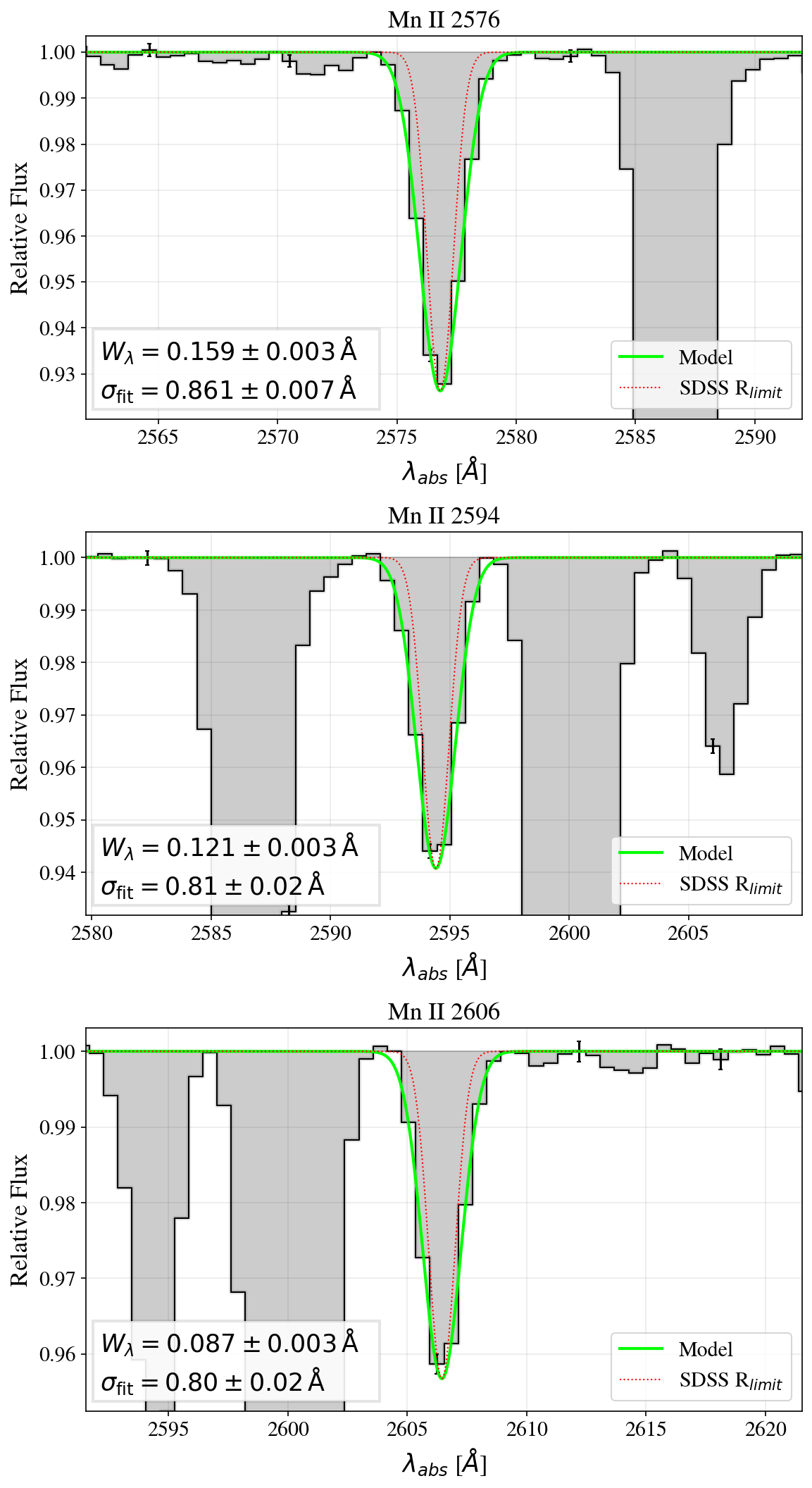}
 \caption{Composites of the three Mn II absorption lines measured, with measured rest-frame equivalent widths shown in the bottom left corner of each subplot.}
 \label{fig:mn_composite}
\end{figure}

\begin{figure*}
 \includegraphics[width=\textwidth]{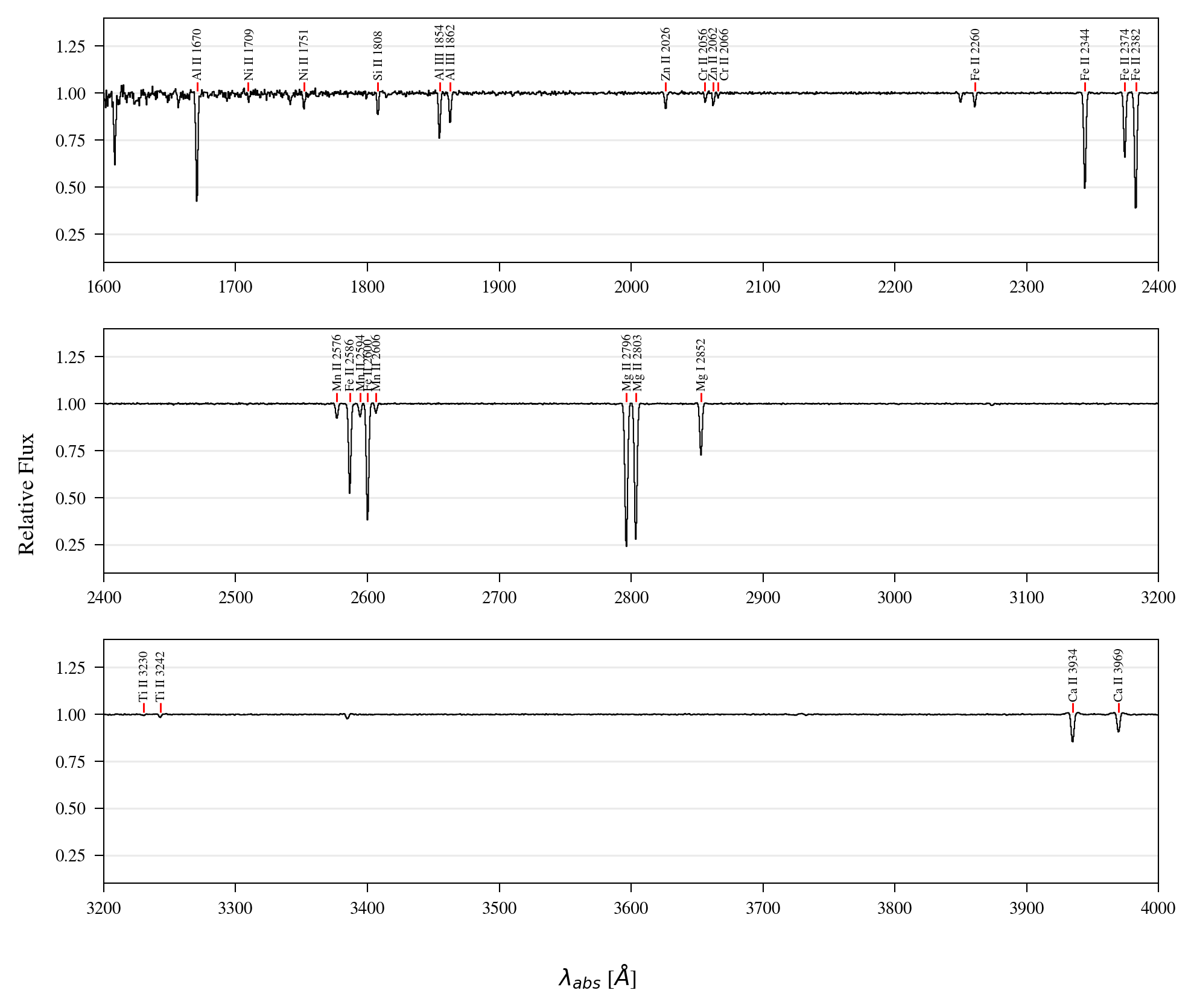}
 \caption{Composite spectra generated from the Ca II absorption catalog of this study. All the lines that had equivalent widths and column densities measured are labelled.}
 \label{fig:composite_paper}
\end{figure*}

To reliably measure the equivalent widths, column densities, and other properties of Ca II absorption lines, we constructed composite spectra from our catalog of 1,646 Ca II absorption detections. Due to the high spectral noise in individual quasar spectra, accurately measuring these features on a per-spectrum basis is challenging. A composite approach allows us to increase the signal-to-noise ratio by stacking many spectra together, thereby enhancing the clarity and fidelity of the absorption features. This technique is essential to produce high-quality data for meaningful measurements, as individual spectra often contain insufficient signal strength for precise analysis of weak metal lines.

The process began with a careful selection of suitable spectra for each line in the composite. We first removed spectra with redshifts that placed the rest-frame wavelength of the line outside the observable spectral range, which would prevent the line from appearing in the spectrum. Additionally, spectra where the absorber frame wavelength fell below 3800 Å, or spectra with lines that were potentially contaminated by absorption features in the Ly$\alpha$ forest, were excluded. This step ensured that the remaining spectra would yield clean, unblended absorption lines, minimizing interference from unrelated spectral features.

After selecting appropriate spectra, we used flux-conserving interpolation to resample each spectrum onto a common wavelength grid to align them for stacking. Each individual spectrum was then continuum-fitted and normalized by dividing out the continuum. This normalization step was crucial for accurately combining the spectra, as it mitigated the effects of variations in quasar continuum shape, allowing for a clearer focus on the absorption lines themselves. The normalized spectra were subsequently stacked using a weighted average, where the inverse square of the flux errors served as weights, giving higher influence to spectra with better SNRs and reducing the impact of noisier data.

One of the challenges in previous studies, such as those by \cite{2024MNRAS.531..387G} and \cite{2023MNRAS.518.5590F}, was the need to manually review each spectrum individually to avoid contamination from large spectral features or blended lines. However, with over a thousand spectra in our sample and 26 separate metal lines to check, this approach was impractical. To address this, we developed an automated threshold-based filtering method described as follows: after stacking the spectra into a preliminary composite for each line, we established a lower-bound threshold, defined as three standard deviations below the central Gaussian profile. Any spectrum with two or more points falling below this threshold within a five-pixel radius of the line center was excluded from the composite. This filtering attempts to remove contamination and spectral features that are clearly not absorption lines, as these can greatly affect the stacked result. To evaluate the effectiveness of this process, we manually examined several weaker lines, such as the Mn II lines. We found that the filtering successfully removed over 80\% of contaminated lines while retaining more than 90\% of valid samples.

Finally, a secondary round of continuum fitting was applied to the composite spectra, as some level of absorption at every pixel lowered the continuum slightly below one. By refitting and centering the continuum around one, we achieved a final composite spectrum that accurately reflected the mean properties of the Ca II absorbers in our sample. Figure \ref{fig:mn_composite} shows example results of the composite fitting of three Mn II absorption lines. 

This robust, multi-step process produced a high-quality composite spectra, setting a solid foundation for measuring equivalent widths, column densities, and other line characteristics with reduced uncertainties. Figure~\ref{fig:composite_paper} illustrates the results of this process, except a larger window of 800 \AA\ is used. In total, we generated composite spectra for 26 metal absorption lines associated with Ca II systems, providing a comprehensive set of data for detailed analysis.

\subsection{2175 \AA\ Dust Absorbers (2DAs)} 
\label{sec:2da}

\setlength{\tabcolsep}{5.1pt} 
\begin{table*}
\centering
\caption{Randomly selected 20 quasars with their fitted parameters and extinction models; the rest are listed in supplementary materials. The table includes the following columns: SDSS jname, redshifts ($z_{\text{qso}}$, $z_{\text{abs}}$), $c_3$ (Drude amplitude), $\gamma$ (Drude FWHM), $A_\text{bump}$ (bump strength), $\Delta A_{\mathrm{bump}}^{\text{MCMC}}$ (fitting error from MCMC), $\Delta A_{\mathrm{bump}}^{\text{FAP}}$ ($1\sigma$ spread of the control samples' $A_\text{bump}$) , $A_V$ (visual extinction), $E(B - V)$ (color excess), Model (best-fitting canonical extinction model), $N_\text{QSO}$ (number of control sample quasars), and $A_V$ median and $\sigma$ (sigma-clipped median and 1$\sigma$ spread of visual extinction values from the control sample).}
\label{tab:2da_params}
\begin{tabular}{lccccccccc|c|ccc}
\hline
SDSS Name & $z_{\mathrm{qso}}$ & $z_{\mathrm{abs}}$ & $c_3$ & $\gamma$ & $A_{\mathrm{bump}}$ & $\Delta A_{\mathrm{bump}}^{\text{MCMC}}$ & $\Delta A_{\mathrm{bump}}^{\text{FAP}}$ & $A_V$ & $E(B-V)$ & Model & \multicolumn{3}{c}{CS $A_V$} \\
 & & & & & & & & & & & $N_{\text{QSO}}$  & Median & $\sigma$ \\
\hline
J001446.79\texttt{+}301718.1 & $1.764$ & $0.876$ & $0.828$ & $0.954$ & $1.365$ & $0.016$ & $0.295$ & $0.200$ & $0.073$ & SMC & 100 & $0.03$ & $0.16$ \\
J003403.59\texttt{-}060334.5 & $1.795$ & $0.875$ & $0.809$ & $1.001$ & $1.270$ & $0.102$ & $0.295$ & $1.420$ & $0.417$ & LMC & 100 & $0.08$ & $0.32$ \\
J004832.05\texttt{+}075630.7 & $1.087$ & $0.877$ & $0.471$ & $0.953$ & $0.777$ & $0.014$ & $0.223$ & $0.136$ & $0.044$ & MW & 100 & $0.04$ & $0.24$ \\
J012342.74\texttt{+}022410.7 & $1.877$ & $0.954$ & $0.514$ & $0.924$ & $0.873$ & $0.026$ & $0.163$ & $0.287$ & $0.084$ & LMC & 100 & $0.09$ & $0.25$ \\
J013526.22\texttt{+}111709.7 & $1.988$ & $1.149$ & $0.288$ & $0.925$ & $0.490$ & $0.007$ & $0.120$ & $0.290$ & $0.106$ & SMC & 87 & $0.04$ & $0.13$ \\
J075500.47\texttt{+}384915.5 & $2.112$ & $1.287$ & $0.205$ & $0.942$ & $0.342$ & $0.011$ & $0.089$ & $0.514$ & $0.186$ & LMC2 & 80 & $0.06$ & $0.28$ \\
J084506.26\texttt{+}214956.6 & $2.501$ & $1.081$ & $0.578$ & $0.948$ & $0.957$ & $0.043$ & $0.189$ & $0.649$ & $0.190$ & LMC & 100 & $-0.03$ & $0.22$ \\
J090122.68\texttt{+}204446.5 & $2.093$ & $1.019$ & $0.583$ & $0.962$ & $0.951$ & $0.010$ & $0.224$ & $0.808$ & $0.237$ & LMC & 100 & $0.03$ & $0.18$ \\
J092438.01\texttt{+}330741.5 & $1.882$ & $0.888$ & $0.697$ & $0.913$ & $1.199$ & $0.078$ & $0.234$ & $0.414$ & $0.133$ & MW & 100 & $0.08$ & $0.26$ \\
J105941.86\texttt{+}351058.1 & $1.793$ & $1.075$ & $0.189$ & $0.935$ & $0.318$ & $0.019$ & $0.083$ & $0.174$ & $0.063$ & LMC2 & 100 & $0.07$ & $0.23$ \\
J110052.14\texttt{+}502044.8 & $1.484$ & $1.060$ & $0.553$ & $0.980$ & $0.886$ & $0.071$ & $0.153$ & $0.346$ & $0.112$ & MW & 100 & $-0.02$ & $0.21$ \\
J110221.55\texttt{+}303918.8 & $1.834$ & $1.331$ & $0.250$ & $0.935$ & $0.420$ & $0.024$ & $0.119$ & $0.442$ & $0.160$ & LMC2 & 100 & $0.03$ & $0.25$ \\
J120113.81\texttt{+}283922.1 & $2.452$ & $1.288$ & $0.378$ & $0.961$ & $0.618$ & $0.036$ & $0.107$ & $0.355$ & $0.129$ & SMC & 100 & $-0.06$ & $0.13$ \\
J121028.93\texttt{+}205544.5 & $2.376$ & $1.325$ & $0.406$ & $0.978$ & $0.652$ & $0.007$ & $0.091$ & $0.577$ & $0.169$ & LMC & 71 & $0.01$ & $0.19$ \\
J162658.32\texttt{+}472716.8 & $1.155$ & $1.104$ & $0.405$ & $0.951$ & $0.669$ & $0.033$ & $0.116$ & $0.886$ & $0.321$ & LMC2 & 100 & $0.02$ & $0.26$ \\
J214717.47\texttt{-}001822.3 & $1.585$ & $1.149$ & $0.191$ & $0.939$ & $0.319$ & $0.015$ & $0.083$ & $0.145$ & $0.047$ & MW & 100 & $0.03$ & $0.14$ \\
J215524.88\texttt{+}185902.0 & $2.152$ & $1.121$ & $0.541$ & $0.935$ & $0.909$ & $0.018$ & $0.106$ & $0.435$ & $0.128$ & LMC & 100 & $-0.03$ & $0.28$ \\
J230120.49\texttt{+}044143.6 & $1.720$ & $1.058$ & $0.324$ & $0.927$ & $0.549$ & $0.011$ & $0.075$ & $0.135$ & $0.049$ & SMC & 100 & $0.03$ & $0.09$ \\
J230721.55\texttt{+}142356.7 & $2.180$ & $1.261$ & $0.452$ & $0.938$ & $0.756$ & $0.041$ & $0.099$ & $0.519$ & $0.189$ & SMC & 100 & $0.01$ & $0.19$ \\
J231036.99\texttt{+}283907.4 & $1.976$ & $1.307$ & $0.271$ & $0.898$ & $0.475$ & $0.021$ & $0.084$ & $0.548$ & $0.200$ & SMC & 100 & $0.05$ & $0.17$ \\
\hline
\end{tabular}

\end{table*}
\begin{table*}
\centering
\caption{Updated extinction values for the 12 2DA detections presented in \citet{2023MNRAS.518.5590F}. Quasar J141951.84+470901.3 did not have enough control samples based on the selection process.}
\label{tab:hfang_2da_params}
\begin{tabular}{lccc|cc|c|cc|cc}
\hline
SDSS Name & $z_{\mathrm{qso}}$ & $z_{\mathrm{abs}}$ & $A_V$ & $E(B-V)$ & Model & $N_{\text{QSO}}$ & \multicolumn{2}{c}{CS $A_V$} & \multicolumn{2}{c}{CS $E(B-V)$} \\
 & & & & & & & Median & $\sigma$ & Median & $\sigma$ \\
\hline
J005251.81\texttt{+}090945.2 & $2.660$ & $1.320$ & $1.20$ & $0.35$ & LMC & 30 & $0.11$ & $0.34$ & $0.03$ & $0.10$\\
    J081738.34\texttt{+}433303.6 & $2.160$ & $1.102$ & $0.97$ & $0.31$ & MW & 100 & $0.03$ & $0.21$ & $0.01$ & $0.07$\\
    J120301.01\texttt{+}063441.5 & $2.175$ & $0.862$ & $0.87$ & $0.26$ & LMC & 100 & $0.02$ & $0.25$ & $0.01$ & $0.07$\\
    J121219.87\texttt{+}293622.8 & $1.392$ & $1.220$ & $0.35$ & $0.13$ & LMC2 & 30 & $-0.12$ & $0.22$ & $-0.04$ & $0.08$\\
    J141850.71\texttt{+}154106.3 & $1.477$ & $1.038$ & $0.31$ & $0.11$ & SMC & 30 & $0.04$ & $0.20$ & $0.01$ & $0.07$\\
    J141951.84\texttt{+}470901.3 & $2.296$ & $1.274$ & $0.51$ & $0.18$ & LMC2 & 3 & - & - & - & -\\
    J154457.53\texttt{+}283325.6 & $1.966$ & $1.070$ & $0.20$ & $0.07$ & SMC & 100 & $-0.01$ & $0.19$ & $0.00$ & $0.07$\\
    J110236.85\texttt{+}235706.5 & $2.119$ & $1.357$ & $0.59$ & $0.21$ & LMC2 & 56 & $0.03$ & $0.24$ & $0.01$ & $0.09$\\
    J135123.47\texttt{+}474712.1 & $1.953$ & $1.063$ & $0.22$ & $0.08$ & SMC & 30 & $0.11$ & $0.24$ & $0.04$ & $0.09$\\
    J082516.44\texttt{+}094700.4 & $2.161$ & $1.290$ & $0.05$ & $0.02$ & LMC & 100 & $0.04$ & $0.20$ & $0.01$ & $0.06$\\
    J100442.90\texttt{+}242226.0 & $1.258$ & $1.110$ & $0.15$ & $0.05$ & SMC & 32 & $-0.09$ & $0.15$ & $-0.03$ & $0.05$\\
    J124914.78\texttt{+}301550.0 & $2.372$ & $1.272$ & $0.72$ & $0.26$ & SMC & 100 & $0.02$ & $0.20$ & $0.01$ & $0.07$\\
\hline
\end{tabular}
\end{table*}

\begin{figure}
 \includegraphics[width=\columnwidth]{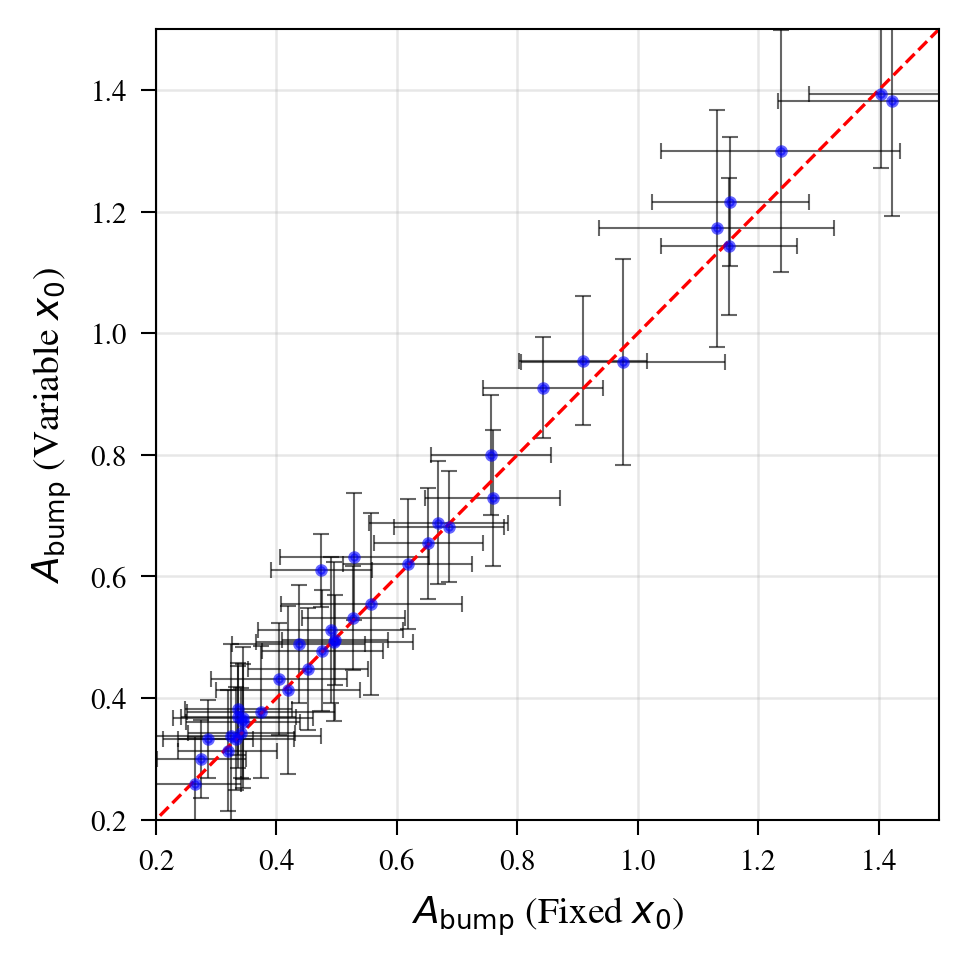}
 \caption{Comparison between $A_\text{bump}$ measurements from fixing $x_0$ and varying $x_0$ for 2DAs with $z_\text{abs} > 1.1$. There is an average 4.6\% difference in bump strength between the two sets of measurements.}
 \label{fig:abump_cmp}
\end{figure}

\begin{figure*}
 \includegraphics[width=\textwidth]{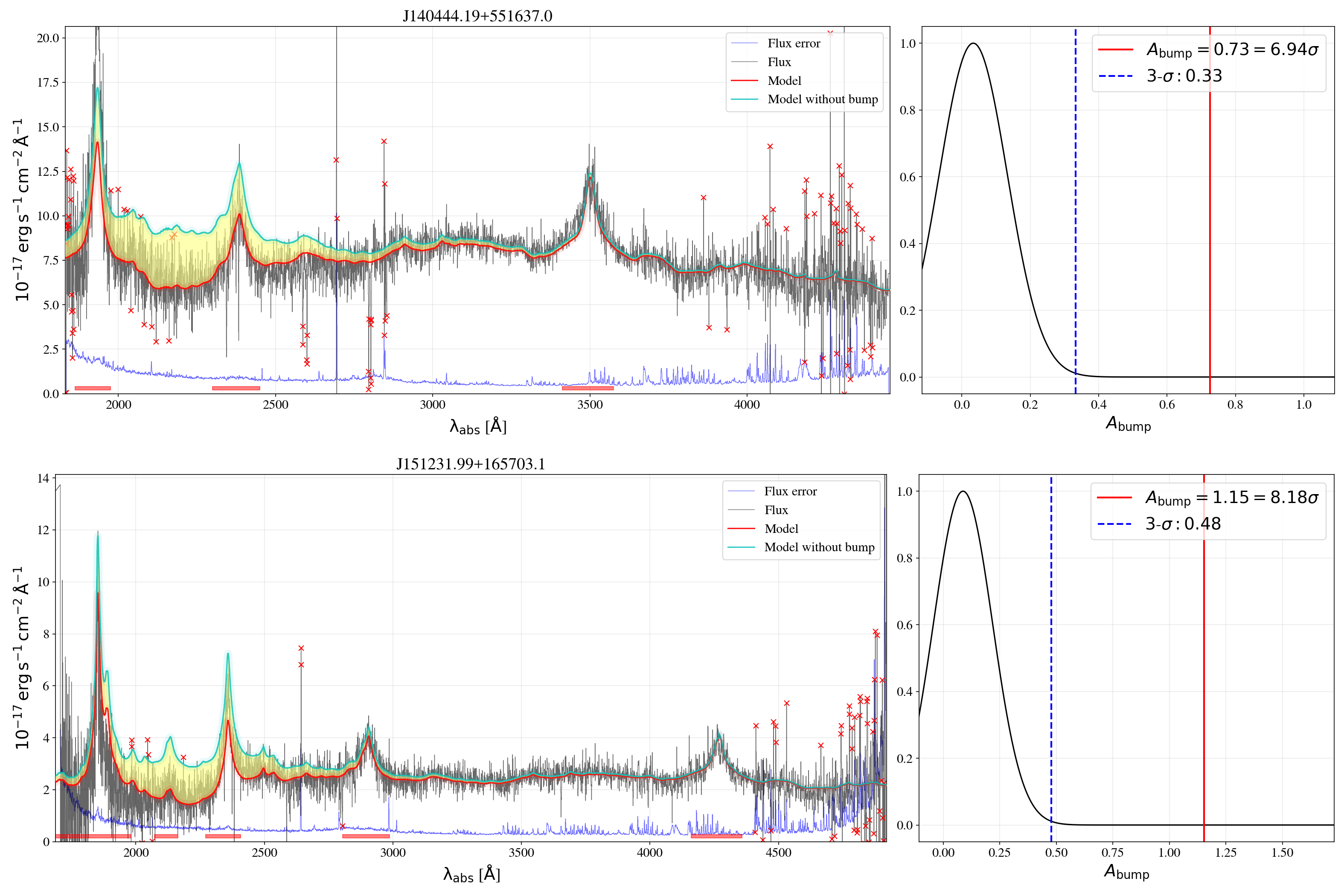}
 \caption{Example of our 2DA bump fitting, adapted from \citet{2011ApJ...732..110J}, for the spectra of quasars J140444.19+551637.0 and J151231.99+165703.1. In the left plots, pink underlined regions are omitted due to their overlap with emission features, while red x's mark data points (outside emission features) that deviate by more than 3 sigma from the quasar continuum. These are also omitted from fitting. The right plots display the bump strength relative to control samples, both of which exceed the 3 sigma (blue dotted line) threshold and are therefore classified as detections.}
 \label{fig:ex_bump_fit}
\end{figure*}

Among the 1,646 Ca II absorbers identified in this study, we analyzed 783 systems with redshifts \(z_{\text{abs}} \geq 0.8\) to search for the presence of the 2175 \AA\ dust bump. The methodology for identifying the bump was adapted from the fitting programs and procedures developed by \citet{2011ApJ...732..110J}, \citet{2023MNRAS.518.5590F}, and \citet{2024MNRAS.531..387G}. 

The fitting process consisted of two main stages. The first stage involved fitting a power-law continuum combined with a Drude profile to measure the bump strength (\(A_{\text{bump}}\)) and assess its significance. Following the approach of \citet{2023MNRAS.518.5590F}, this stage included two key steps: (1) fitting the parameters \(c_1\), \(c_2\), and \(c_3\) while keeping \(\gamma\) fixed at 0.945, which is the mean value characteristic of LMC2 extinction \citep{2003ApJ...594..279G}. This was necessary because \(\gamma\) is degenerate with \(c_1\) and \(c_2\), and most absorbers were estimated to have LMC-like extinction; (2) fitting the parameter \(\gamma\) to complete the model. To ensure fitting accuracy, data points in quasar regions close to major emission features were omitted from the fitting in both steps. Additionally, a global continuum was fitted to each spectrum, and data points that deviated by more than 3 sigma from the continuum were also omitted to minimize the influence spectral features such as absorption lines. In cases where broad absorption lines (BALs) were present, a custom masking procedure was implemented to ensure that these regions were not considered during the fitting process.

A key difference in our approach compared to previous studies (\citealt{2024MNRAS.531..387G}; \citealt{2023MNRAS.518.5590F}) was that we fixed the peak position of the bump, \(x_0\), to its theoretical value of 4.598 throughout the entire process. This decision was based on our assessment that many of our low-redshift samples have incomplete bumps, and allowing \(x_0\) to vary introduces significant uncertainties and lead to bogus fitting results. In cases where $z_\text{abs}>1.1$, which meant almost the entirety of the bump was within SDSS spectral range, we tested the same procedure with varying $x_0$, and found an average 5\% differences in $A_\text{bump}$ measurements (Figure ~\ref{fig:abump_cmp}).

To determine the statistical significance of the detected bumps, we used False-Alarm Probability (FAP) maps from \citet{zhao2020thesis}, matching the measured \(\gamma\), \(z_{\text{abs}}\), and \(z_{\text{qso}}\) values with the control sample distribution of \(A_{\text{bump}}\) derived from non-absorber quasars. Those absorbers with bump strengths exceeding the 3\(\sigma\) threshold compared to the control sample distributions were classified as detections. 

Following this procedure, we identified 95 of the 783 absorbers as 2DAs. These detections include 57 (22.2\%) among 257 strong absorbers and 38 (7.2\%) among 526 weak absorbers. Examples of 2DA detections are shown in Figure~\ref{fig:ex_bump_fit}.


In our second stage, we fitted known extinction models (MW, LMC, LMC2, and SMC) to determine the total extinction value ($A_V$) for each system. The best-fitting canonical model for each absorber was selected based on minimizing the residuals. This model was then used to calculate an estimate values for visual extinction $A_V$ and colour excess $E(B-V)$.

To evaluate the significance of our extinction measurements, we constructed control samples following the methodology outlined in \citet{2024MNRAS.531..387G}. For each of the 95 2DA detections, we chose up to 100 control samples by drawing reference spectra from the SDSS quasar catalog, excluding all Mg II absorbers. Reference spectra were selected to match the absorber spectrum in emission redshift ($z_{\text{em}}$) and the $z$-band magnitude, using thresholds of $|\Delta z_{\text{em}}| \leq 0.1$ and $|\Delta z| \le 0.2$. In cases where fewer than 30 control samples were available, we incrementally increased the threshold on $|\Delta z|$ by 0.01 until at least 30 control samples were obtained, following the methodology from \cite{2015A&A...580A...8L} and \cite{2024MNRAS.531..387G}.

The extinction fitting procedure was then applied to these control samples, and the resulting median and standard deviation of $A_V$ were measured for each 2DA detection. The parameters derived for a random 20 of our 2DA detections are presented in Table~\ref{tab:2da_params}.

\citet{2023MNRAS.518.5590F} reported that 33\% (7 / 21) of their strong Ca II absorbers exhibited the 2175 Å dust bump, while 6\% (5 / 79) of their weak absorbers did. The discrepancy between our percentage (22\%) for strong absorbers and theirs is likely due to two reasons:
(1) We fixed $x_0$, the peak bump position, during our fitting process, whereas \citet{2023MNRAS.518.5590F} allowed it to vary. This could have introduced minor differences, particularly for lower redshift targets.
(2) Our sample size for strong absorbers is significantly larger, with over 12 times more data compared to their study. As a result, our findings likely reflect a more statistically robust representation of the strong Ca II absorber population. 

Additionally, Table 4 of \citet{2023MNRAS.518.5590F} contained several incorrect values in the $A_V$ column. In this work, we present an updated version of the extinction values (Table~\ref{tab:hfang_2da_params}), which includes corrected values based on the approach described in this study.

\subsection{The Curve of Growth}

\begin{figure}
 \includegraphics[width=\columnwidth]{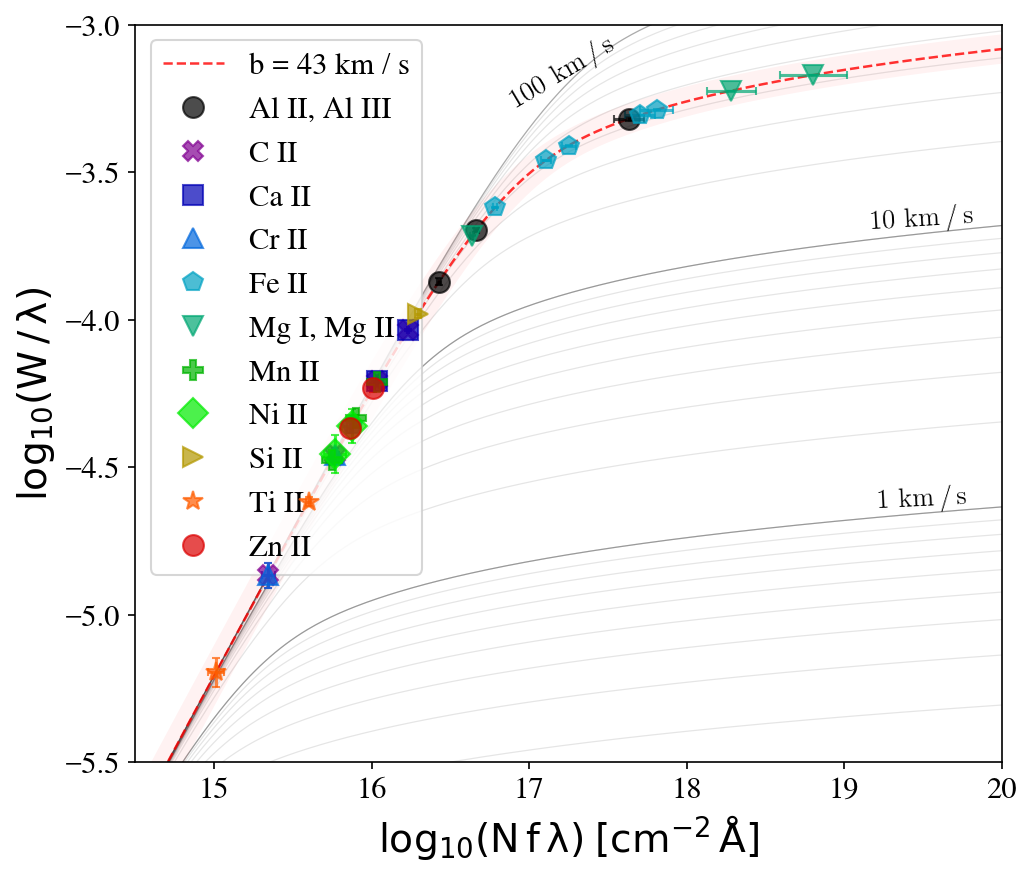}
 \caption{A curve of growth (COG) constructed from our 1,646 Ca II absorbers. A best fit of five Fe II generates the curve, and the resulting Doppler parameter is measured to be $b = 43 \pm 2 \,\text{km  s}^{-1}$.} \label{fig:final_cog}
\end{figure}

\begin{table}
    \centering
    \caption{Comparison of Doppler broadening parameter ($b$) for various absorbers across studies. The $b$ parameter is calculated from constructed COGs.}
    \label{tab:doppler_broad_comparison}
    \begin{tabular}{lccr} 
        \hline
         & Cumulative & Strong & Weak \\
        \hline
        Ca II (Fang et al.) & - & $63.6$ & $46.3$\\
        Ca II (This study) & $43\pm 2$ & $58\pm2$ & $42\pm3$\\
        2DA (This study) & $45 \pm 3$& - & - \\
        \hline
    \end{tabular}
\end{table}

\begin{figure}
 \includegraphics[width=\columnwidth]{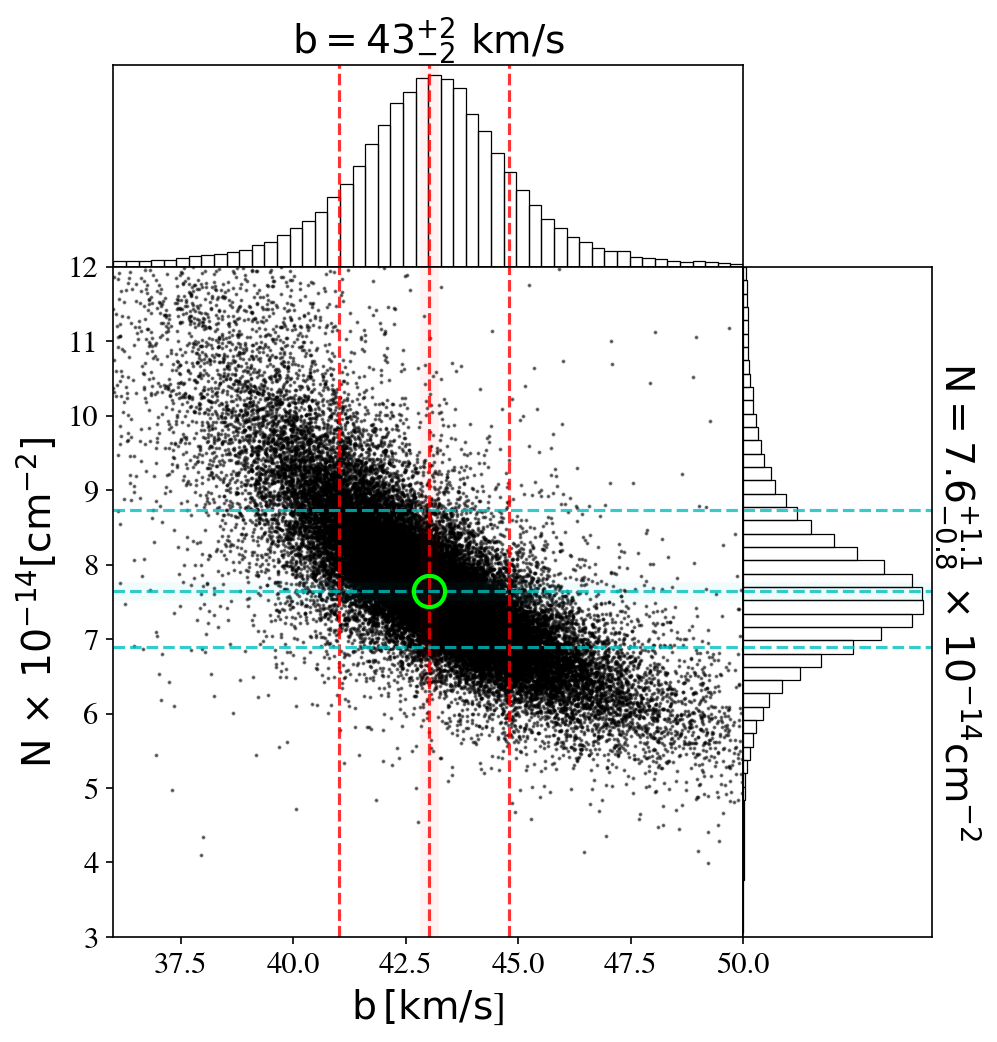}
 \caption{MCMC corner diagram which shows the sampling of each parameter ($N$ and $b$).}
 \label{fig:final_cog_corner_diagram}
\end{figure}

To determine the column densities of various elements in the composite spectrum, we constructed a curve of growth (COG), a diagnostic tool widely used to analyze absorption systems. The COG plots $\log_{10}(N f \lambda)$ against $\log_{10}(W / \lambda)$, where \(N\) is the column density, \(f\) is the oscillator strength from \cite{2014MNRAS.438..388M}, \(\lambda\) is the rest wavelength of that specific absorption line, and \(W\) is the equivalent width. By relating these quantities, the COG provides insight into gas kinematics, Doppler broadening, line saturation, and abundances of different metals. Figure~\ref{fig:final_cog} presents the resulting COG based on our composite spectrum.

To construct our COG, we utilized the program created by \cite{2024MNRAS.531..387G}, which calculates the best fit of the measurements of six Fe II absorption lines, namely Fe II 2260, Fe II 2344, Fe II 2374, Fe II 2382, Fe II 2586, and Fe II 2600. To estimate the physical parameters underlying the COG, a Markov chain Monte Carlo (MCMC) method is implemented. This produced 5,000 samples that follow and estimate the distribution of $N$ and $b$ from the six Fe II lines. The resulting Doppler broadening parameter, b, was found to be $b = 43 \pm 2 \,\text{ km  s}^{-1}$, and the corner diagram from the MCMC sampling is shown in Figure~\ref{fig:final_cog_corner_diagram}.

To further investigate our catalog, we followed \cite{2023MNRAS.518.5590F} by dividing our Ca II sample into two strong and weak absorbers. For each subgroup, we generated independent composite measurements and constructed their corresponding COGs. We measured $b=42\pm3\,\text{ km  s}^{-1}$ for the weak subcatalog and $b=58\pm2\,\text{ km  s}^{-1}$ for the strong subcatalog. Similar to \cite{2023MNRAS.518.5590F}, the Doppler parameter for the strong absorbers was found to be significantly larger than the parameter for the weak absorbers. Our measured values, along with those reported by \cite{2023MNRAS.518.5590F} are presented in Table~\ref{tab:doppler_broad_comparison}. The $b$ parameters of our strong and weak subcatalogs are comparable to those of \cite{2023MNRAS.518.5590F}, with both within 2$\sigma$ of each other. The slight differences is likely due to us using the most recent $f$ values from \cite{2014MNRAS.438..388M} while \cite{2023MNRAS.518.5590F} using $f$ values from \cite{1991ApJS...77..119M}.

\subsection{Column Densities and Relative Element Abundances}

\begin{table}
	\centering
        \caption{Properties and measurements of the fitted absorption lines. The rest-frame wavelength ($\lambda_0$) values are taken from \citet{1988ApJS...68..449M}, and the oscillator strengths ($f$) are sourced from \citet{2014MNRAS.438..388M}. The rest-frame equivalent widths ($\mathrm{W}_0$) are derived from our fitted spectra, while the column densities ($\mathrm{N}$) are calculated using Equation~\eqref{eq:column_density}. Uncertainties are provided where applicable.}
	\label{tab:column_densities}
	\begin{tabular}{lcccr} 
        \hline
        Species & $\mathrm{\lambda}_0$ & $f$ & $\mathrm{W}_0$ & $\mathrm{N}$ \\
        \hline
        Al II & 1670.79 & 1.7400 & 0.80 ± 0.01 & $\geq 1.9 \times 10^{13}$ \\
        \hline
        Al III & 1854.72 & 0.5590 & 0.374 ± 0.006 & $\geq 2.2 \times 10^{13}$ \\
        Al III & 1862.79 & 0.2780 & 0.251 ± 0.006 & $\geq 2.9 \times 10^{13}$ \\
        \hline
        Ca II & 3934.78 & 0.6267 & 0.364 ± 0.003 & $4.24 (\pm 0.04 ) \times 10^{12}$ \\
        Ca II & 3969.59 & 0.3116 & 0.246 ± 0.003 & $5.65 (\pm 0.07 ) \times 10^{12}$ \\
        \hline
        Cr II & 2056.25 & 0.1030 & 0.072 ± 0.004 & $1.9 (\pm 0.1 ) \times 10^{13}$ \\
        Cr II & 2066.16 & 0.0512 & 0.028 ± 0.003 & $1.5 (\pm 0.1 ) \times 10^{13}$ \\
        \hline
        Fe II & 2260.78 & 0.0024 & 0.100 ± 0.003 & $9.0 (\pm 0.2 ) \times 10^{14}$ \\
        Fe II & 2344.21 & 0.1140 & 0.914 ± 0.003 & $\geq 1.6 \times 10^{14}$ \\
        Fe II & 2374.46 & 0.0313 & 0.573 ± 0.003 & $\geq 3.7 \times 10^{14}$ \\
        Fe II & 2382.77 & 0.3200 & 1.229 ± 0.003 & $\geq 7.6 \times 10^{13}$ \\
        Fe II & 2586.65 & 0.0691 & 0.900 ± 0.003 & $\geq 2.2 \times 10^{14}$ \\
        Fe II & 2600.17 & 0.2390 & 1.288 ± 0.003 & $\geq 9.0 \times 10^{13}$ \\
        \hline
        Mg I & 2852.96 & 1.8300 & 0.551 ± 0.003 & $4.18 (\pm 0.02 ) \times 10^{12}$ \\
        \hline
        Mg II & 2796.35 & 0.6155 & 1.894 ± 0.003 & $\geq 4.4 \times 10^{13}$ \\
        Mg II & 2803.53 & 0.3058 & 1.678 ± 0.003 & $\geq 7.9 \times 10^{13}$ \\
        \hline
        Mn II & 2576.88 & 0.3610 & 0.159 ± 0.003 & $7.5 (\pm 0.1 ) \times 10^{12}$ \\
        Mn II & 2594.50 & 0.2800 & 0.121 ± 0.003 & $7.2 (\pm 0.2 ) \times 10^{12}$ \\
        Mn II & 2606.46 & 0.1980 & 0.087 ± 0.003 & $7.3 (\pm 0.2 ) \times 10^{12}$ \\
        \hline
        Ni II & 1709.60 & 0.0324 & 0.0750 ± 0.01 & $9 (\pm 1 ) \times 10^{13}$ \\
        Ni II & 1751.91 & 0.0277 & 0.061 ± 0.009 & $8 (\pm 1 ) \times 10^{13}$ \\
        \hline
        Si II & 1808.01 & 0.0021 & 0.190 ± 0.007 & $3.2 (\pm 0.1 ) \times 10^{15}$ \\
        \hline
        Ti II & 3073.88 & 0.1210 & 0.020 ± 0.002 & $1.9 (\pm 0.2 ) \times 10^{12}$ \\
        Ti II & 3384.74 & 0.3580 & 0.082 ± 0.003 & $2.25 (\pm 0.08 ) \times 10^{12}$ \\
        \hline
        Zn II & 2026.14 & 0.5010 & 0.119 ± 0.004 & $6.5 (\pm 0.2 ) \times 10^{12}$ \\
        Zn II & 2062.66 & 0.2460 & 0.089 ± 0.004 & $9.6 (\pm 0.4 ) \times 10^{12}$ \\
        \hline
	\end{tabular}
\end{table}

\begin{table}
    \centering
    \caption{Relative abundance measurements for various elements relative to Zn in Ca II absorbers. Values are provided for all Ca II absorbers, strong Ca II absorbers, and weak Ca II absorbers.}
    \label{tab:x_zn_measurements}
    \begin{tabular}{lccc}
        \hline
        X & [X/Zn] (All) & [X/Zn] (Strong) & [X/Zn] (Weak) \\
        \hline
        Si II & $-0.31 \pm 0.02$ & $-0.74 \pm 0.06$ & $-0.26 \pm 0.02$ \\
        Mn II & $-0.86 \pm 0.01$ & $-0.89 \pm 0.02$ & $-0.87 \pm 0.01$ \\
        Cr II & $-0.67 \pm 0.03$ & $-0.77 \pm 0.06$ & $-0.61 \pm 0.03$ \\
        Fe II & $-0.84 \pm 0.02$ & $-0.92 \pm 0.03$ & $-0.81 \pm 0.02$ \\
        Ni II & $-0.58 \pm 0.04$ & $-0.53 \pm 0.07$ & $-0.62 \pm 0.05$ \\
        Ti II & $-0.90 \pm 0.02$ & $-1.09 \pm 0.03$ & $-0.93 \pm 0.02$ \\
        \hline
    \end{tabular}
\end{table}

From the COG, we could differentiate between the satuated and unsaturated absorption lines. The column densities ($N$) of the unsaturated lines, ones that were located on the linear portion of the curve, were calculated with the Equation~\eqref{eq:column_density}.

\begin{equation}
    N\approx1.13\times10^{20}\frac{W}{f\lambda^2}\text{cm}^{-2}.
    \label{eq:column_density}
\end{equation}

Table~\ref{tab:column_densities} lists the elements on which we performed spectral fitting, along with their measured equivalent widths and column densities derived using Equation~\eqref{eq:column_density}.

Relative abundances of elements compared to the Sun, using Zn II as a metallicity indicator, were calculated with Equation~\eqref{eq:relative_abundance}:

\begin{equation}
    \left[\frac{\text{X}}{\text{Zn}}\right] = \log_{10} \left[\frac{N(\text{X})}{N(\text{Zn})}\right] - \log_{10} \left[\frac{N(\text{X})_{\text{sun}}}{N(\text{Zn})_{\text{sun}}}\right],
    \label{eq:relative_abundance}
\end{equation}

where $N(\text{X})$ represents the average column densities of element X's unsaturated absorption lines, and $N(\text{X})_{\text{sun}}$ denotes solar abundances as reported by \cite{2009ARA&A..47..481A}. 

In addition to the overall Ca II absorber catalog, we computed relative abundances separately for the strong and weak subsets of the catalog. All calculated abundances are presented in Table~\ref{tab:x_zn_measurements}.

\begin{figure}
 \includegraphics[width=\columnwidth]{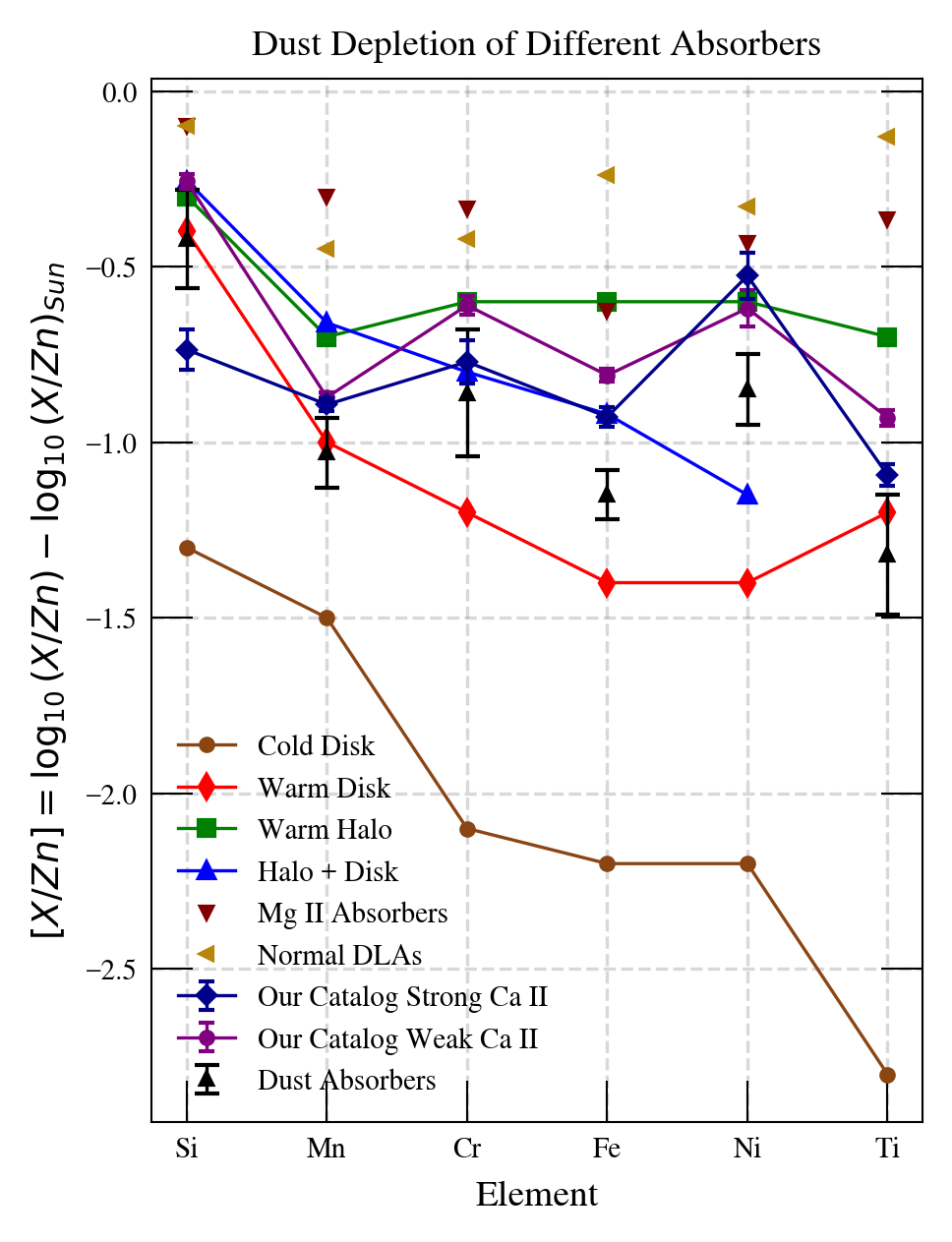}
 \caption{Relative abundances of various elements compared to Zn II, plotted for our strong Ca II absorbers and our weak Ca II absorbers. These results are compared to cold disk gas, warm disk gas, warm halo gas, and halo + disk gas \citep{1996ARA&A..34..279S, 1999ApJ...512..636W}; Mg II absorber relative element abundances \citep{2006MNRAS.367..945Y} normal DLA relative abundances \citep{2016MNRAS.458.4074Q}; and 2175 \AA\ dust absorber relative abundances \citep{2018MNRAS.474.4870M}. }
 \label{fig:relative_abundances}
\end{figure}

Following the methodology of \cite{2023MNRAS.518.5590F} and \cite{2024MNRAS.531..387G}, we plotted the relative abundances against reference data sets, including measurements for cold disk gas, warm disk gas, warm halo gas, and a combination of halo and disk gas \citep{1996ARA&A..34..279S, 1999ApJ...512..636W}. Additional comparisons were made to Mg II absorbers \citep{2009ARA&A..47..481A}, normal DLA systems \citep{2016MNRAS.458.4074Q}, strong and weak Ca II absorbers \citep{2023MNRAS.518.5590F}, and 2175 \AA\ dust absorbers \citep{2018MNRAS.474.4870M}. These comparisons are visualized in Figure~\ref{fig:relative_abundances}. 

Overall, our results are consistent with those reported by \citet{2014MNRAS.444.1747S} and \citet{2023MNRAS.518.5590F}. Specifically, weak absorbers in our sample exhibit depletion patterns resembling those of the Warm Halo component, while strong absorbers show greater similarity to the combined Halo + disk component. The main outlier is the Ni II depletion in strong absorbers, which is likely due to the limited number of strong systems in our catalog with redshifts high enough to cover the Ni II lines. Even among these few cases, the Ni II lines often fall near the blue edge of the spectra, where data quality is poor and contamination is heavy. Additionally, the abundance estimates for the systems compared in the plot are based on $f$-values drawn from different studies, which also contributes to discrepancies across all the data points.

\subsection{Reddening}

\begin{figure}
 \includegraphics[width=\columnwidth]{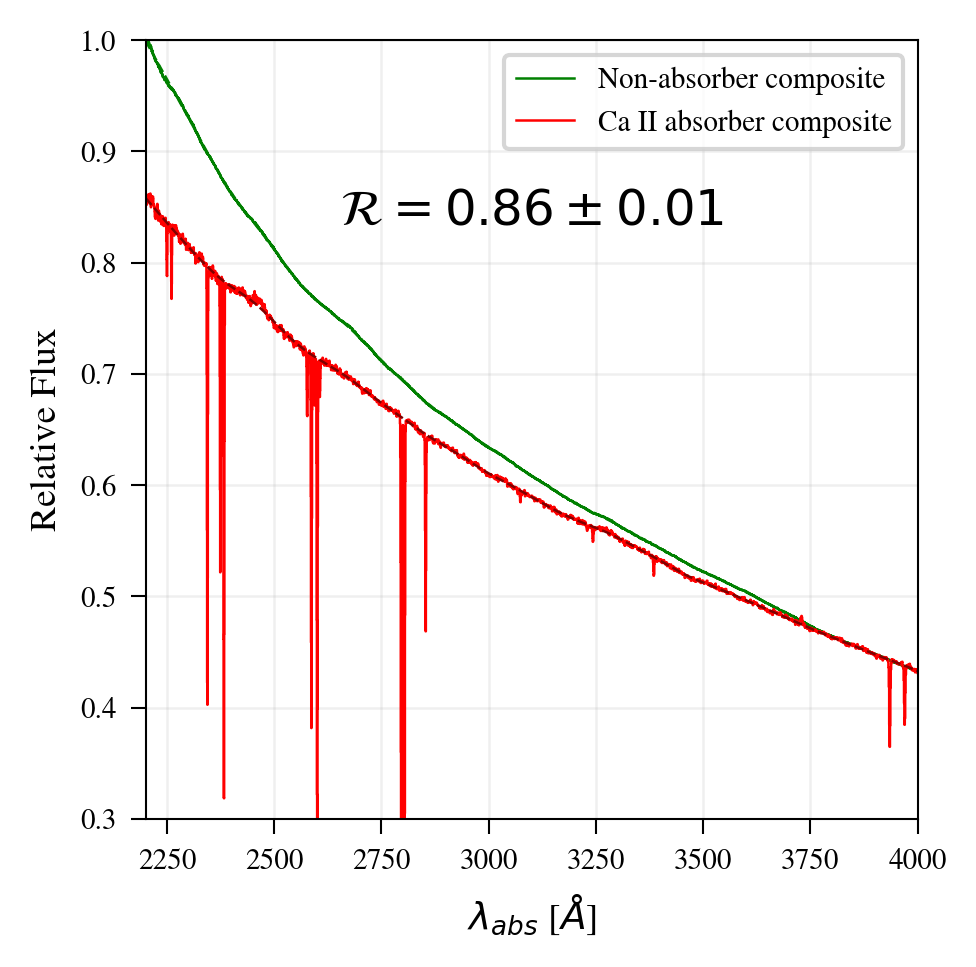}
 \caption{A comparison of the full Ca II absorber detection catalog composite and control samples of the absorbers. The $\mathcal{R}$ parameter is a measure of the absorber reddening, defined as the flux ratio between the two composites at a rest wavelength of 2200 \AA.}
 \label{fig:ca_reddening_composite}
\end{figure}

\begin{figure}
 \includegraphics[width=\columnwidth]{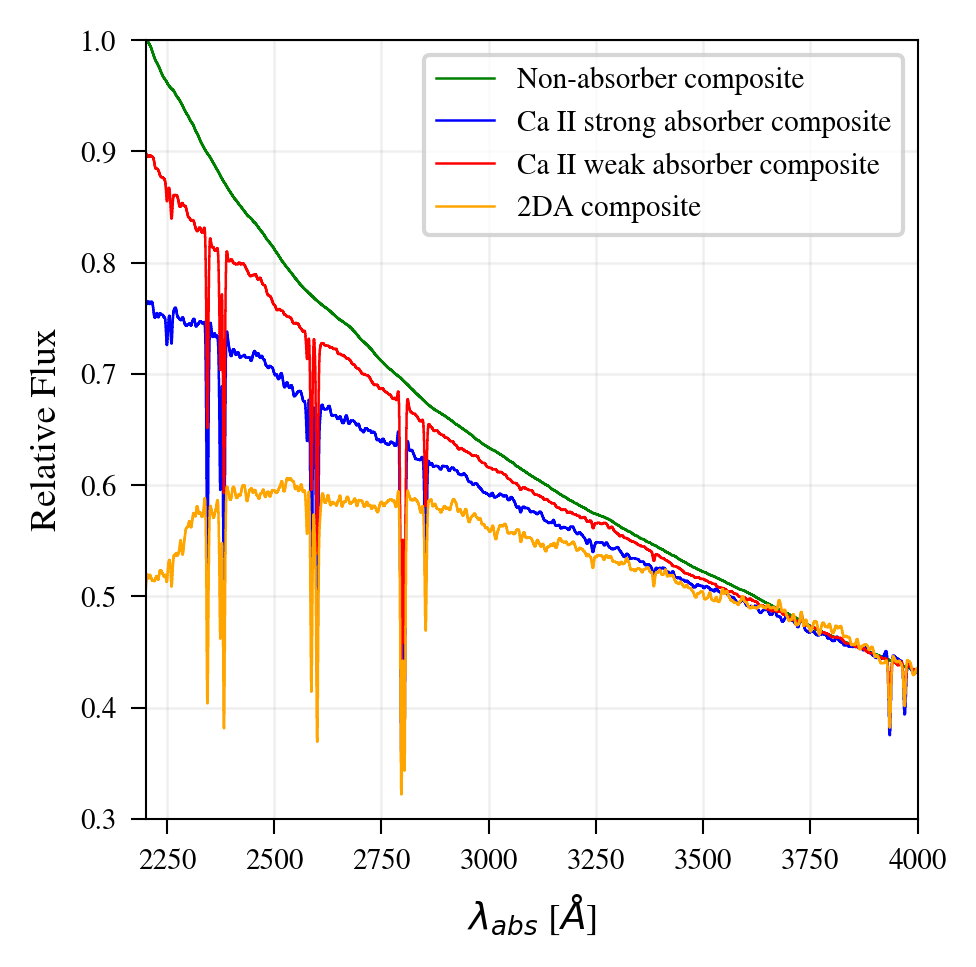}
 \caption{A comparison between the composites from our strong (N = 520) and weak (N = 1126) Ca II absorption catalogs and the composite of our 2DA absorbers. The green curve shows the non-absorber control sample composite used in this study.}
 \label{fig:ca_reddening_composite_comparison}
\end{figure}

To determine the average reddening associated with Ca II absorbers, we followed a methodology similar to that of previous studies, particularly \cite{2023MNRAS.518.5590F} and \cite{2024MNRAS.531..387G}. This method compares absorber spectra to matched, unabsorbed reference samples using composite spectra to quantify reddening effects. Specifically,
we created a control sample for each Ca II absorber in our catalog,
drawing samples from our search catalog after removing all
Mg II absorbers. Then, reference spectra were selected to match the
Ca II absorber spectra in terms of $z$-band magnitude and emission
redshift ($z_\text{em}$), with thresholds set to $|\Delta z| \le 0.2$ and $|z_\text{em}| \le 0.1$.

For each Ca II absorber in our catalog, up to 400 control spectra were selected, shifted to the rest frame, and interpolated onto a single wavelength grid spanning from 2200 \AA\ to 4500 \AA. By constructing a composite from the references, we obtain a baseline continuum representative of unreddened quasar spectra which do not have absorbers.

To quantify reddening, we used the parameter $\mathcal{R}$, which is defined as the flux ratio between the absorber and non-absorber composites at a rest wavelength of 2200 \AA. The error in $\mathcal{R}$ was estimated with the method established by \cite{2023MNRAS.518.5590F}: The RMS of each composite spectrum within a 100\AA\ window centered at 2200\AA\ was used to estimate the continuum uncertainty, which was then propagated through the ratio calculation to obtain the error on $\mathcal{R}$.

As seen in Figure~\ref{fig:ca_reddening_composite}, our analysis yielded a reddening measurement of $\mathcal{R} = 0.86 \pm 0.01$, showing a clear, statistically significant attenuation due to dust associated with Ca II absorbers. 

To investigate the reddening characteristics further, we again divided our catalog into two subsets based on line strength. For each subgroup, we generated independent composites and calculated separate $\mathcal{R}$ values. The results indicate distinct reddening levels: strong absorbers showed an $\mathcal{R}$ value of $0.76 \pm 0.01$, while weak absorbers exhibited a slightly higher $\mathcal{R}$ of $0.88 \pm 0.01$. We also generated a composite for our 2DAs, and found that they had an $\mathcal{R}$ value of $0.51\pm 0.01$, the lowest of the all. Both our strong absorbers and dust absorbers had similar values to what \citet{2023MNRAS.518.5590F} reported.

Our catalog contains roughly ten times the number of weak absorbers compared to \cite{2023MNRAS.518.5590F}, potentially influencing our lower measured $\mathcal{R}$ value. To test this possibility, we randomly divided our 1126 weak absorbers into 10 subsets, each containing approximately 100 absorbers—similar in size to their sample. The resulting $\mathcal{R}$ values were 0.92, 0.89, 0.85, 0.86, 0.81, 0.93, 0.81, 0.84, 0.91, and 0.84, yielding a mean of $0.89$ and a standard deviation of $0.04$. Compared to the reported value of $0.97 \pm 0.02$, these results differ by $1.8\sigma$, indicating reasonable agreement. Additionally, running the original weak subcatalog from \cite{2023MNRAS.518.5590F} through our program yielded $\mathcal{R}=0.95\pm 0.02$, which is slightly lower than their published result—likely due to minor differences in the selected control samples.

In addition to our main analysis, we conducted a test to further investigate the impact of reddening in different absorber populations. We selected 1,646 Mg II absorbers from the DR16 Mg II catalog \citep{2021MNRAS.504...65A} that do not contain any Ca II absorption and have a similar redshift distribution to our Ca II absorber catalog. For this sample, we measured $\mathcal{R} = 0.99 \pm 0.01$, indicating almost no reddening compared to the Ca II absorbers. Figure~\ref{fig:ca_reddening_composite_comparison} plots the strong and weak Ca II composites, as well as the dust absorber composite against the control samples. 

Table~\ref{tab:reddening_comparison} compares our reddening measurements with those from previous studies, including results from \cite{2015MNRAS.452.3192S}, \cite{2023MNRAS.518.5590F}, and \cite{2024MNRAS.531..387G}. All of our Ca II absorber and dust absorber measurement values are comparable to those from prior studies. These results further confirm the presence of a reddening effect linked to Ca II absorption and support the interpretation that Ca II absorbers are indicative of dusty galactic environments.

\begin{table}
	\centering
	\caption{Comparison of reddening parameters ($\mathcal{R}$) for Ca II, Mg II, C I, and 2175 dust absorbers across studies. The $\mathcal{R}$ parameter is calculated from the flux ratio between absorbers and non-absorbers at a rest-frame wavelength of 2200 Å.}
	\label{tab:reddening_comparison}
	\begin{tabular}{lccr} 
		\hline
	     & Cumulative & Strong & Weak \\
		\hline
		Ca II (Sardane et al.) & 0.85 & 0.73 & 0.95\\
		Ca II (Fang et al.) & $0.92\pm 0.02$ & $0.76\pm 0.05$ & $0.97\pm 0.02$\\
		Ca II (This study) & $0.86\pm 0.01$ & $0.77\pm 0.01$ & $0.89 \pm 0.01$\\
            2DA (Fang et al.) & $0.53\pm 0.05$ & - & -\\
            2DA (This study) & $0.51\pm 0.01$ & - & -\\
            Mg II (This study) & $0.99\pm 0.01$ & - & -\\
		C I (Ge et al.) & $0.89\pm 0.01$ & - & -\\
            \hline
	\end{tabular}
\end{table}

\subsection{Absorber Occurrence  Rates}

\begin{table}
	\centering
	\caption{Counts of strong and weak Ca II absorbers, all Ca II absorbers, Mg II absorbers, and dust absorbers across different absorption redshift ranges.}
	\label{tab:abs_counts}
	\begin{tabular}{lccccr} 
		\hline
	      $z_\text{abs}$ Range & Ca II Strong & Ca II Weak & Ca II & Mg II & 2DA \\
		\hline
                (0.35, 1.4) & 520 & 1126 & 1646 & 108783 & - \\
                (0.35, 0.8) & 263 & 600 & 863 & 39738 & - \\
                (0.8, 1.4) & 257 & 526 & 783 & 69045 & 95 \\
            \hline
	\end{tabular}
\end{table}

With our expanded catalogs of Ca II and dust absorbers, we can now compute more robust estimates of their occurrence rates within Mg II absorber populations. Table~\ref{tab:abs_counts} summarizes the counts of Ca II absorbers—categorized by strong and weak—as well as Mg II absorbers and identified 2DAs, split into two redshift intervals. We find that roughly 1.5\% of Mg II absorbers host Ca II absorption. Among Ca II systems at $0.8 < z_{\text{abs}} < 1.4$, about 12\% exhibit the 2175 \AA\ dust feature, corresponding to 22\% of strong absorbers and 7\% of weak absorbers.

\section{Conclusions and Discussions}
\label{sec:Conclusions_and_Discussions}

This study presents a significant advancement in the study of Ca II absorbers by employing a dual CNN-based methodology, resulting in a vetted catalogue of 1,646 Ca II absorbers, including 1,121 new systems. This work expands the largest existing catalog of Ca II absorbers by over threefold, providing a robust data set for further scientific exploration. The innovative use of a dual CNN approach, combining Ca II detection with Fe II cross-verification, proved crucial in achieving high detection accuracy while minimizing false positives. 

Curve of growth analysis shows similar results to \cite{2023MNRAS.518.5590F}, with strong Ca II having a significantly higher Doppler broadening parameter compared to weak absorbers, thus indicating more velocity components in the strong absorber environments. Our results on dust depletion also align with previous studies: strong Ca II systems exhibit signatures consistent with Milky Way disk+halo environments, while weak systems more closely resemble halo environments. Further, we find reddening in Ca II systems consistent with prior studies, with one exception: our weak Ca II sample appears slightly dustier than previously reported. Future work includes Mg II–independent searches to recover low-redshift Ca II systems ($z_\text{abs} < 0.36$) and obtain unbiased statistics. These methods can also be applied to the Dark Energy Spectroscopic Instrument (DESI) Survey, which provides much more quasar spectra for analysis.

\section*{Acknowledgements}

Funding for the Sloan Digital Sky Survey IV has been provided by the Alfred P. Sloan Foundation, the U.S. Department of Energy Office of Science, and the Participating Institutions. SDSS acknowledges support and resources from the Center for High-Performance Computing at the University of Utah. The SDSS web site is www.sdss4.org.

SDSS is managed by the Astrophysical Research Consortium for the Participating Institutions of the SDSS Collaboration including the Brazilian Participation Group, the Carnegie Institution for Science, Carnegie Mellon University, Center for Astrophysics | Harvard \& Smithsonian (CfA), the Chilean Participation Group, the French Participation Group, Instituto de Astrofísica de Canarias, The Johns Hopkins University, Kavli Institute for the Physics and Mathematics of the Universe (IPMU) / University of Tokyo, the Korean Participation Group, Lawrence Berkeley National Laboratory, Leibniz Institut für Astrophysik Potsdam (AIP), Max-Planck-Institut für Astronomie (MPIA Heidelberg), Max-Planck-Institut für Astrophysik (MPA Garching), Max-Planck-Institut für Extraterrestrische Physik (MPE), National Astronomical Observatories of China, New Mexico State University, New York University, University of Notre Dame, Observatório Nacional / MCTI, The Ohio State University, Pennsylvania State University, Shanghai Astronomical Observatory, United Kingdom Participation Group, Universidad Nacional Autónoma de México, University of Arizona, University of Colorado Boulder, University of Oxford, University of Portsmouth, University of Utah, University of Virginia, University of Washington, University of Wisconsin, Vanderbilt University, and Yale University.

\section*{Data Availability}

The data underlying this article were accessed from the Sloan Digital Sky Survey (\href{https://www.sdss4.org/dr16/}{https://www.sdss4.org/dr16/}). The data generated from this research are available in the article and in its online supplementary material.



\bibliographystyle{mnras}
\bibliography{references} 

\begin{thebibliography}{}
\makeatletter
\relax
\def\mn@urlcharsother{\let\do\@makeother \do\$\do\&\do\#\do\^\do\_\do\%\do\~}
\def\mn@doi{\begingroup\mn@urlcharsother \@ifnextchar [ {\mn@doi@} {\mn@doi@[]}}
\def\mn@doi@[#1]#2{\def\@tempa{#1}\ifx\@tempa\@empty \href {http://dx.doi.org/#2} {doi:#2}\else \href {http://dx.doi.org/#2} {#1}\fi \endgroup}
\def\mn@eprint#1#2{\mn@eprint@#1:#2::\@nil}
\def\mn@eprint@arXiv#1{\href {http://arxiv.org/abs/#1} {{\tt arXiv:#1}}}
\def\mn@eprint@dblp#1{\href {http://dblp.uni-trier.de/rec/bibtex/#1.xml} {dblp:#1}}
\def\mn@eprint@#1:#2:#3:#4\@nil{\def\@tempa {#1}\def\@tempb {#2}\def\@tempc {#3}\ifx \@tempc \@empty \let \@tempc \@tempb \let \@tempb \@tempa \fi \ifx \@tempb \@empty \def\@tempb {arXiv}\fi \@ifundefined {mn@eprint@\@tempb}{\@tempb:\@tempc}{\expandafter \expandafter \csname mn@eprint@\@tempb\endcsname \expandafter{\@tempc}}}

\bibitem[\protect\citeauthoryear{{Anand}, {Nelson}  \& {Kauffmann}}{{Anand} et~al.}{2021}]{2021MNRAS.504...65A}
{Anand} A.,  {Nelson} D.,   {Kauffmann} G.,  2021, \mn@doi [\mnras] {10.1093/mnras/stab871}, \href {https://ui.adsabs.harvard.edu/abs/2021MNRAS.504...65A} {504, 65}

\bibitem[\protect\citeauthoryear{{Asplund}, {Grevesse}, {Sauval}  \& {Scott}}{{Asplund} et~al.}{2009}]{2009ARA&A..47..481A}
{Asplund} M.,  {Grevesse} N.,  {Sauval} A.~J.,   {Scott} P.,  2009, \mn@doi [\araa] {10.1146/annurev.astro.46.060407.145222}, \href {https://ui.adsabs.harvard.edu/abs/2009ARA&A..47..481A} {47, 481}

\bibitem[\protect\citeauthoryear{{Fang}, {Xia}, {Ge}, {Willis}  \& {Zhao}}{{Fang} et~al.}{2023}]{2023MNRAS.518.5590F}
{Fang} H.,  {Xia} I.,  {Ge} J.,  {Willis} K.,   {Zhao} Y.,  2023, \mn@doi [\mnras] {10.1093/mnras/stac3473}, \href {https://ui.adsabs.harvard.edu/abs/2023MNRAS.518.5590F} {518, 5590}

\bibitem[\protect\citeauthoryear{{Ge}, {Willis}, {Chao}, {Jan}, {Zhao}  \& {Fang}}{{Ge} et~al.}{2024}]{2024MNRAS.531..387G}
{Ge} J.,  {Willis} K.,  {Chao} K.,  {Jan} A.,  {Zhao} Y.,   {Fang} H.,  2024, \mn@doi [\mnras] {10.1093/mnras/stae799}, \href {https://ui.adsabs.harvard.edu/abs/2024MNRAS.531..387G} {531, 387}

\bibitem[\protect\citeauthoryear{{Gordon}, {Clayton}, {Misselt}, {Landolt}  \& {Wolff}}{{Gordon} et~al.}{2003}]{2003ApJ...594..279G}
{Gordon} K.~D.,  {Clayton} G.~C.,  {Misselt} K.~A.,  {Landolt} A.~U.,   {Wolff} M.~J.,  2003, \mn@doi [\apj] {10.1086/376774}, \href {https://ui.adsabs.harvard.edu/abs/2003ApJ...594..279G} {594, 279}

\bibitem[\protect\citeauthoryear{{Jiang}, {Ge}, {Zhou}, {Wang}  \& {Wang}}{{Jiang} et~al.}{2011}]{2011ApJ...732..110J}
{Jiang} P.,  {Ge} J.,  {Zhou} H.,  {Wang} J.,   {Wang} T.,  2011, \mn@doi [\apj] {10.1088/0004-637X/732/2/110}, \href {https://ui.adsabs.harvard.edu/abs/2011ApJ...732..110J} {732, 110}

\bibitem[\protect\citeauthoryear{Kingma \& Ba}{Kingma \& Ba}{2017}]{kingma2017adammethodstochasticoptimization}
Kingma D.~P.,  Ba J.,  2017, Adam: A Method for Stochastic Optimization

\bibitem[\protect\citeauthoryear{Kramida, Ralchenko, Reader  \& Team}{Kramida et~al.}{2024}]{NIST_ASD}
Kramida A.,  Ralchenko Y.,  Reader J.,   Team N.~A.,  2024, NIST Atomic Spectra Database (version 5.12), \url{https://physics.nist.gov/asd}, \mn@doi{10.18434/T4W30F}

\bibitem[\protect\citeauthoryear{{Ledoux}, {Noterdaeme}, {Petitjean}  \& {Srianand}}{{Ledoux} et~al.}{2015}]{2015A&A...580A...8L}
{Ledoux} C.,  {Noterdaeme} P.,  {Petitjean} P.,   {Srianand} R.,  2015, \mn@doi [\aap] {10.1051/0004-6361/201424122}, \href {https://ui.adsabs.harvard.edu/abs/2015A&A...580A...8L} {580, A8}

\bibitem[\protect\citeauthoryear{{Liu}, {Li}, {Gao}, {Zhang}, {Xu}, {Wang}  \& {Lin}}{{Liu} et~al.}{2025}]{2025ApJS..276...37L}
{Liu} Y.,  {Li} J.,  {Gao} L.,  {Zhang} H.,  {Xu} Z.,  {Wang} Y.,   {Lin} W.,  2025, \mn@doi [\apjs] {10.3847/1538-4365/ad9250}, \href {https://ui.adsabs.harvard.edu/abs/2025ApJS..276...37L} {276, 37}

\bibitem[\protect\citeauthoryear{{Lyke} et~al.,}{{Lyke} et~al.}{2020}]{2020ApJS..250....8L}
{Lyke} B.~W.,  et~al., 2020, \mn@doi [\apjs] {10.3847/1538-4365/aba623}, \href {https://ui.adsabs.harvard.edu/abs/2020ApJS..250....8L} {250, 8}

\bibitem[\protect\citeauthoryear{{Ma} et~al.,}{{Ma} et~al.}{2018}]{2018MNRAS.474.4870M}
{Ma} J.,  et~al., 2018, \mn@doi [\mnras] {10.1093/mnras/stx3123}, \href {https://ui.adsabs.harvard.edu/abs/2018MNRAS.474.4870M} {474, 4870}

\bibitem[\protect\citeauthoryear{{Morton}}{{Morton}}{1991}]{1991ApJS...77..119M}
{Morton} D.~C.,  1991, \mn@doi [\apjs] {10.1086/191601}, \href {https://ui.adsabs.harvard.edu/abs/1991ApJS...77..119M} {77, 119}

\bibitem[\protect\citeauthoryear{{Morton}, {York}  \& {Jenkins}}{{Morton} et~al.}{1988}]{1988ApJS...68..449M}
{Morton} D.~C.,  {York} D.~G.,   {Jenkins} E.~B.,  1988, \mn@doi [\apjs] {10.1086/191295}, \href {https://ui.adsabs.harvard.edu/abs/1988ApJS...68..449M} {68, 449}

\bibitem[\protect\citeauthoryear{{Murphy} \& {Berengut}}{{Murphy} \& {Berengut}}{2014}]{2014MNRAS.438..388M}
{Murphy} M.~T.,  {Berengut} J.~C.,  2014, \mn@doi [\mnras] {10.1093/mnras/stt2204}, \href {https://ui.adsabs.harvard.edu/abs/2014MNRAS.438..388M} {438, 388}

\bibitem[\protect\citeauthoryear{{Nestor}, {Turnshek}  \& {Rao}}{{Nestor} et~al.}{2005}]{2005ApJ...628..637N}
{Nestor} D.~B.,  {Turnshek} D.~A.,   {Rao} S.~M.,  2005, \mn@doi [\apj] {10.1086/427547}, \href {https://ui.adsabs.harvard.edu/abs/2005ApJ...628..637N} {628, 637}

\bibitem[\protect\citeauthoryear{{Nestor}, {Pettini}, {Hewett}, {Rao}  \& {Wild}}{{Nestor} et~al.}{2008}]{2008MNRAS.390.1670N}
{Nestor} D.~B.,  {Pettini} M.,  {Hewett} P.~C.,  {Rao} S.,   {Wild} V.,  2008, \mn@doi [\mnras] {10.1111/j.1365-2966.2008.13857.x}, \href {https://ui.adsabs.harvard.edu/abs/2008MNRAS.390.1670N} {390, 1670}

\bibitem[\protect\citeauthoryear{{Petitjean}, {Srianand}  \& {Ledoux}}{{Petitjean} et~al.}{2000}]{2000A&A...364L..26P}
{Petitjean} P.,  {Srianand} R.,   {Ledoux} C.,  2000, \mn@doi [\aap] {10.48550/arXiv.astro-ph/0011437}, \href {https://ui.adsabs.harvard.edu/abs/2000A&A...364L..26P} {364, L26}

\bibitem[\protect\citeauthoryear{{Quider}, {Nestor}, {Turnshek}, {Rao}, {Monier}, {Weyant}  \& {Busche}}{{Quider} et~al.}{2011}]{2011AJ....141..137Q}
{Quider} A.~M.,  {Nestor} D.~B.,  {Turnshek} D.~A.,  {Rao} S.~M.,  {Monier} E.~M.,  {Weyant} A.~N.,   {Busche} J.~R.,  2011, \mn@doi [\aj] {10.1088/0004-6256/141/4/137}, \href {https://ui.adsabs.harvard.edu/abs/2011AJ....141..137Q} {141, 137}

\bibitem[\protect\citeauthoryear{{Quiret} et~al.,}{{Quiret} et~al.}{2016}]{2016MNRAS.458.4074Q}
{Quiret} S.,  et~al., 2016, \mn@doi [\mnras] {10.1093/mnras/stw524}, \href {https://ui.adsabs.harvard.edu/abs/2016MNRAS.458.4074Q} {458, 4074}

\bibitem[\protect\citeauthoryear{Rimoldini}{Rimoldini}{2007}]{Rimoldini2007}
Rimoldini L.~G.,  2007, PhD thesis, University of Pittsburgh

\bibitem[\protect\citeauthoryear{{Sardane}, {Turnshek}  \& {Rao}}{{Sardane} et~al.}{2014}]{2014MNRAS.444.1747S}
{Sardane} G.~M.,  {Turnshek} D.~A.,   {Rao} S.~M.,  2014, \mn@doi [\mnras] {10.1093/mnras/stu1554}, \href {https://ui.adsabs.harvard.edu/abs/2014MNRAS.444.1747S} {444, 1747}

\bibitem[\protect\citeauthoryear{{Sardane}, {Turnshek}  \& {Rao}}{{Sardane} et~al.}{2015}]{2015MNRAS.452.3192S}
{Sardane} G.~M.,  {Turnshek} D.~A.,   {Rao} S.~M.,  2015, \mn@doi [\mnras] {10.1093/mnras/stv1506}, \href {https://ui.adsabs.harvard.edu/abs/2015MNRAS.452.3192S} {452, 3192}

\bibitem[\protect\citeauthoryear{{Savage} \& {Sembach}}{{Savage} \& {Sembach}}{1996}]{1996ARA&A..34..279S}
{Savage} B.~D.,  {Sembach} K.~R.,  1996, \mn@doi [\araa] {10.1146/annurev.astro.34.1.279}, \href {https://ui.adsabs.harvard.edu/abs/1996ARA&A..34..279S} {34, 279}

\bibitem[\protect\citeauthoryear{{Welty}, {Frisch}, {Sonneborn}  \& {York}}{{Welty} et~al.}{1999}]{1999ApJ...512..636W}
{Welty} D.~E.,  {Frisch} P.~C.,  {Sonneborn} G.,   {York} D.~G.,  1999, \mn@doi [\apj] {10.1086/306795}, \href {https://ui.adsabs.harvard.edu/abs/1999ApJ...512..636W} {512, 636}

\bibitem[\protect\citeauthoryear{{Wild} \& {Hewett}}{{Wild} \& {Hewett}}{2005}]{2005MNRAS.361L..30W}
{Wild} V.,  {Hewett} P.~C.,  2005, \mn@doi [\mnras] {10.1111/j.1745-3933.2005.00058.x}, \href {https://ui.adsabs.harvard.edu/abs/2005MNRAS.361L..30W} {361, L30}

\bibitem[\protect\citeauthoryear{{Wild}, {Hewett}  \& {Pettini}}{{Wild} et~al.}{2006}]{2006MNRAS.367..211W}
{Wild} V.,  {Hewett} P.~C.,   {Pettini} M.,  2006, \mn@doi [\mnras] {10.1111/j.1365-2966.2005.09935.x}, \href {https://ui.adsabs.harvard.edu/abs/2006MNRAS.367..211W} {367, 211}

\bibitem[\protect\citeauthoryear{{Wild}, {Hewett}  \& {Pettini}}{{Wild} et~al.}{2007}]{2007MNRAS.374..292W}
{Wild} V.,  {Hewett} P.~C.,   {Pettini} M.,  2007, \mn@doi [\mnras] {10.1111/j.1365-2966.2006.11146.x}, \href {https://ui.adsabs.harvard.edu/abs/2007MNRAS.374..292W} {374, 292}

\bibitem[\protect\citeauthoryear{{Wolfe}, {Turnshek}, {Smith}  \& {Cohen}}{{Wolfe} et~al.}{1986}]{1986ApJS...61..249W}
{Wolfe} A.~M.,  {Turnshek} D.~A.,  {Smith} H.~E.,   {Cohen} R.~D.,  1986, \mn@doi [\apjs] {10.1086/191114}, \href {https://ui.adsabs.harvard.edu/abs/1986ApJS...61..249W} {61, 249}

\bibitem[\protect\citeauthoryear{{Xia}, {Ge}, {Willis}  \& {Zhao}}{{Xia} et~al.}{2022}]{2022MNRAS.517.4902X}
{Xia} I.,  {Ge} J.,  {Willis} K.,   {Zhao} Y.,  2022, \mn@doi [\mnras] {10.1093/mnras/stac2905}, \href {https://ui.adsabs.harvard.edu/abs/2022MNRAS.517.4902X} {517, 4902}

\bibitem[\protect\citeauthoryear{{York} et~al.,}{{York} et~al.}{2000}]{2000AJ....120.1579Y}
{York} D.~G.,  et~al., 2000, \mn@doi [\aj] {10.1086/301513}, \href {https://ui.adsabs.harvard.edu/abs/2000AJ....120.1579Y} {120, 1579}

\bibitem[\protect\citeauthoryear{{York} et~al.,}{{York} et~al.}{2006}]{2006MNRAS.367..945Y}
{York} D.~G.,  et~al., 2006, \mn@doi [\mnras] {10.1111/j.1365-2966.2005.10018.x}, \href {https://ui.adsabs.harvard.edu/abs/2006MNRAS.367..945Y} {367, 945}

\bibitem[\protect\citeauthoryear{{Zhao}}{{Zhao}}{2020}]{zhao2020thesis}
{Zhao} Y.,  2020, PhD thesis, University of Florida

\bibitem[\protect\citeauthoryear{{Zhao}, {Ge}, {Yuan}, {Zhao}, {Wang}  \& {Li}}{{Zhao} et~al.}{2019}]{2019MNRAS.487..801Z}
{Zhao} Y.,  {Ge} J.,  {Yuan} X.,  {Zhao} T.,  {Wang} C.,   {Li} X.,  2019, \mn@doi [\mnras] {10.1093/mnras/stz1197}, \href {https://ui.adsabs.harvard.edu/abs/2019MNRAS.487..801Z} {487, 801}

\bibitem[\protect\citeauthoryear{{Zych}, {Murphy}, {Pettini}, {Hewett}, {Ryan-Weber}  \& {Ellison}}{{Zych} et~al.}{2007}]{2007MNRAS.379.1409Z}
{Zych} B.~J.,  {Murphy} M.~T.,  {Pettini} M.,  {Hewett} P.~C.,  {Ryan-Weber} E.~V.,   {Ellison} S.~L.,  2007, \mn@doi [\mnras] {10.1111/j.1365-2966.2007.12015.x}, \href {https://ui.adsabs.harvard.edu/abs/2007MNRAS.379.1409Z} {379, 1409}

\makeatother
\end{thebibliography}




\appendix




\bsp	
\label{lastpage}
\end{document}